\documentclass[12pt]{article}
\usepackage{amssymb}

\def\hybrid{\topmargin 0pt      \oddsidemargin 0pt
        \headheight 0pt \headsep 0pt
        \voffset=-0.5cm
        \hoffset=-0.25in
        \textwidth 6.75in
        \textheight 9.5in       
        \marginparwidth 0.0in
        \parskip 5pt plus 1pt   \jot = 1.5ex}
\catcode`\@=11
\def\marginnote#1{}

\newcount\hour
\newcount\minute
\newtoks\amorpm
\hour=\time\divide\hour by60 \minute=\time{\multiply\hour by60 \global\advance\minute by-\hour}
\edef\standardtime{{\ifnum\hour<12 \global\amorpm={am}%
        \else\global\amorpm={pm}\advance\hour by-12 \fi
        \ifnum\hour=0 \hour=12 \fi
        \number\hour:\ifnum\minute<10 0\fi\number\minute\the\amorpm}}
\edef\militarytime{\number\hour:\ifnum\minute<10 0\fi\number\minute}

\def\draftlabel#1{{\@bsphack\if@filesw {\let\thepage\relax
   \xdef\@gtempa{\write\@auxout{\string
      \newlabel{#1}{{\@currentlabel}{\thepage}}}}}\@gtempa
   \if@nobreak \ifvmode\nobreak\fi\fi\fi\@esphack}
        \gdef\@eqnlabel{#1}}
\def\@eqnlabel{}
\def\@vacuum{}
\def\draftmarginnote#1{\marginpar{\raggedright\scriptsize\tt#1}}
\def\draftlabel#1{{\@bsphack\if@filesw {\let\thepage\relax
   \xdef\@gtempa{\write\@auxout{\string
      \newlabel{#1}{{\@currentlabel}{\thepage}}}}}\@gtempa
   \if@nobreak \ifvmode\nobreak\fi\fi\fi\@esphack}
        \gdef\@eqnlabel{#1}}
\def\@eqnlabel{}
\def\@vacuum{}
\def\draftmarginnote#1{\marginpar{\raggedright\scriptsize\tt#1}}

\def\draft{\oddsidemargin -.5truein
        \def\@oddfoot{\sl preliminary draft \hfil
        \rm\thepage\hfil\sl\today\quad\militarytime}
        \let\@evenfoot\@oddfoot \overfullrule 3pt
        \let\label=\draftlabel
        \let\marginnote=\draftmarginnote
   \def\@eqnnum{(\theequation)\rlap{\kern\marginparsep\tt\@eqnlabel}%
\global\let\@eqnlabel\@vacuum}  }

\def\numberbysection{\@addtoreset{equation}{section}
        \def\theequation{\thesection.\arabic{equation}}}

\def\underline#1{\relax\ifmmode\@@underline#1\else
        $\@@underline{\hbox{#1}}$\relax\fi}

\def\titlepage{\@restonecolfalse\if@twocolumn\@restonecoltrue\onecolumn
     \else \newpage \fi \thispagestyle{empty}\c@page\z@
        \def\thefootnote{\fnsymbol{footnote}} }

\def\endtitlepage{\if@restonecol\twocolumn \else  \fi
        \def\thefootnote{\arabic{footnote}}
        \setcounter{footnote}{0}}  

\numberbysection \hybrid

\newcounter{mo}

\newcommand{\tr}{{\rm tr}}
\newcommand{\ti}[1]{\tilde{#1}}

\newcommand{\Si}{\Sigma}

\newcommand{\ad}{{\rm ad}}

\newcommand{\Om}{\Omega}
\newcommand{\de}{\delta}

\newcommand{\te}{\theta}

\newcommand{\La}{\Lambda}

\newcommand{\ve}{\varepsilon}
\newcommand{\ep}{\epsilon}

\newcommand{\si}{\sigma}

\def\bff{{\bf f}}

\def\bfp{{\bf p}}

\def\cB{{\cal B}}

\def\cE{{\cal E}}
\def\cH{{\cal H}}

\def\cA{{\cal A}}
\def\cG{{\cal G}}

\def\cP{{\cal P}}
\def\cR{{\cal R}}
\def\cT{{\cal T}}

\def\mC{{\mathbb C}}
\def\mZ{{\mathbb Z}}
\def\mR{{\mathbb R}}

\def\frak{\mathfrak}

\def\gM{{\frak M}}

\newcommand{\la}{\lambda}
\newcommand{\vf}{\varphi}
\newcommand{\al}{\alpha}
\newcommand{\be}{\beta}
\newcommand{\ga}{\gamma}
\newcommand{\om}{\omega}
\newcommand{\vth}{\vartheta}

\newcommand{\bfq}{{\bf{q}}}
\newcommand{\bfx}{{\bf{x}}}

\newcommand{\GLN}{{\rm GL}(N,{\mathbb C})}

\newcommand{\gln}{{\rm gl}(N, {\mathbb C})}

\newcommand{\SLZ}{{\rm SL}_2( {\mathbb Z})}
\def\f1#1{\frac{1}{#1}}
\def\oh{\frac{1}2}

\def\bA{\bar A}
\def\bp{\bar \partial}
\def\bz{\bar z}

\newtheorem{predl}{Proposition}[section]

\def\beq{\begin{equation}}
\def\eq{\end{equation}}
\def\p{\partial}

\newtheorem{theor}{Theorem}

\newcommand{\mat}[4]{\left(\begin{array}{cc}{#1}&{#2}\\ \ \\{#3}&{#4}
\end{array}\right)}

\def\res{\mathop{\hbox{Res}}\limits}

\begin{document}

\setcounter{page}{1}

\date{}
\date{}
\vspace{50mm}

\begin{flushright}
 ITEP-TH-12/14\\
\end{flushright}
\vspace{3mm}

\begin{center}
{\LARGE{ Relativistic Classical Integrable Tops}}
 \\ \vspace{2mm} {\LARGE{ and Quantum R-matrices} }
\\
\vspace{12mm} {\large \ \ \ \ {A. Levin}\,$^{\flat\,\sharp}$ \ \ \ \
{M. Olshanetsky}\,$^{\sharp\,\natural}$
\ \ \ \ {A. Zotov}\,$^{\diamondsuit\, \sharp\, \natural}$ }\\
 \vspace{10mm}

 \vspace{2mm} $^\flat$ -- {\small{\sf 
 NRU HSE, Department of Mathematics,
 Myasnitskaya str. 20,  Moscow,  101000,  Russia}}\\
 \vspace{2mm} $^\sharp$ -- {\small{\sf 
 ITEP, B. Cheremushkinskaya str. 25,  Moscow, 117218, Russia}}\\
 \vspace{2mm} $^\natural$ -- {\small{\sf MIPT, Inststitutskii per.  9, Dolgoprudny,
 Moscow region, 141700, Russia}}\\
\vspace{2mm} $^\diamondsuit$ -- {\small{\sf Steklov Mathematical
Institute  RAS, Gubkina str. 8, Moscow, 119991,  Russia}}\\
\end{center}

\begin{center}\footnotesize{{\rm E-mails:}{\rm\ \
 alevin@hse.ru,\  olshanet@itep.ru,\  zotov@mi.ras.ru}}\end{center}

 \begin{abstract}
We describe classical top-like integrable systems arising from the
quantum exchange relations and corresponding Sklyanin algebras. The
Lax operator is expressed in terms of the quantum non-dynamical
$R$-matrix even at the classical level, where the Planck constant
plays the role of the relativistic deformation parameter in the
sense of Ruijsenaars and Schneider (RS). The integrable systems
(relativistic tops) are described as multidimensional Euler tops,
and the inertia tensors are written in terms of the quantum and
classical $R$-matrices. A particular case of ${\rm gl}_N$ system is
gauge equivalent to the $N$-particle RS model while a generic top is
related to the spin generalization of the RS model. The simple
relation between quantum $R$-matrices and classical Lax operators is
exploited in two ways. In the elliptic case we use the Belavin's
quantum $R$-matrix to describe the relativistic classical tops. Also
by the passage to the noncommutative torus we study the large $N$
limit corresponding to the relativistic version of the nonlocal 2d
elliptic hydrodynamics. Conversely, in the rational case we obtain a
new ${\rm gl}_N$ quantum rational non-dynamical $R$-matrix via the
relativistic top, which we get in a different way -- using the
factorized form of the RS Lax operator and the classical Symplectic
Hecke (gauge) transformation. In particular case of ${\rm gl}_2$ the
quantum rational $R$-matrix is the 11-vertex. It was previously
found by Cherednik. At last, we describe the integrable spin chains
and Gaudin models related to the obtained $R$-matrix.
 \end{abstract}

\newpage

{\small{

\tableofcontents

}}

\section{Introduction}
\setcounter{equation}{0}

We start with the quantum exchange relations \cite{Sklyanin0} for
the quantum ${\rm gl}_N$-valued $L$-operators:
 \beq\label{vv001}
 \begin{array}{c}
  \displaystyle{
 R^\hbar_{12}(z-w)\,{\hat L^\eta}_1(z)\,{\hat L^\eta}_2(w)={\hat L^\eta}_2(w)\,{\hat
 L^\eta}_1(z)\,R^\hbar_{12}(z-w)\,,
 }
 \end{array}
 \eq
where the quantum non-dynamical $R$-matrix satisfies the quantum
Yang-Baxter equation
 \beq\label{vv002}
 \begin{array}{c}
  \displaystyle{
 R^\hbar_{12}(z-w)\,R^\hbar_{13}(z)\,R^\hbar_{23}(w)=R^\hbar_{23}(w)\,
 R^\hbar_{13}(z)\, R^\hbar_{12}(z-w)
 }
 \end{array}
 \eq
 and unitarity condition
  \beq\label{vv00271}
 \begin{array}{c}
  \displaystyle{
 R^\hbar_{12}(z)\,R^\hbar_{21}(-z)= f^\hbar(z)\,1\otimes 1
 }
 \end{array}
 \eq
with some function $f^\hbar(z)$.

In this paper we consider a class of solutions of (\ref{vv001}) and
(\ref{vv002}) having simple pole at $z=0$ and satisfying relation
\beq\label{vv7025}
  \begin{array}{c}
  \displaystyle{
 {\hat L^\eta}(z)=\tr_2 \left(R^{\,\eta}_{12}(z)\hat S_2\right)\,,\
 \ \hat S=\res\limits_{z=0} {\hat L}^\eta(z)\,,
 }
 \end{array}
 \eq
 where $\hat S$ is ${\rm gl}_N$-valued operator.
Then (\ref{vv001}) leads to (quadratic) Sklyanin algebra
\cite{Sklyanin} for $\hat S$ which we denote as ${\mathcal
A}_{\hbar,\eta}^{\hbox{\tiny{Skl}}}$. Notice here that we use two
parameters $\hbar$ and $\eta$ in (\ref{vv001}) (it is customary to
consider $\eta=\hbar$). In fact, one can even eliminate the
$\eta$-dependence (see (\ref{vv7116})) but we will see that it is
useful to keep two free parameters from the very beginning.

In the classical limit $\hbar\rightarrow 0$ the matrix components of
the residue $\hat S$ become $\mathbb C$-valued coordinates on the
phase space of an integrable system described by the Lax matrix
$L^\eta(z)$ (it coincides with $\hat L^\eta(z)$, where $\hat S$ is
replaced with ${\rm gl}(N,\mathbb C)$-valued $S$)
 \beq\label{vv7026}
  \begin{array}{|c|}
  \hline\\
  \displaystyle{
 { L^\eta}(z)=\tr_2 \left(R^{\,\eta}_{12}(z) S_2\right)\,,\
 \  S=\res\limits_{z=0}  L^\eta(z)
 }
  \\ \ \\
 \hline
 \end{array}
 \eq
and the standard quadratic $r$-matrix structure: 
 \beq\label{vv704}
 \begin{array}{c}
  \displaystyle{
\{L^\eta_1(z)\,, L^\eta_2(w)\}=[ L^\eta_1(z)\,
L^\eta_2(w),r_{12}(z-w)]
 }
 \end{array}
 \eq
 with the classical $r$-matrix $r_{12}(z)$.
We call this type of models  {\em relativistic integrable tops}
because $\eta$ will be shown to play the role of the relativistic
deformation parameter in the sense of Ruijsenaars and Schneider
\cite{Ruijs1}. The underlying Poincar\'e invariance is discussed in
Section \ref{Poincare}.

 Thus, when $\eta=\hbar$ we have simple relation
(\ref{vv7026}) between the classical Lax operator and the quantum
$R$-matrix, i.e. having quantum $R$-matrix we can define the
classical integrable system.
Write $R$-matrix in the standard ${\rm gl}_N$ basis $\left(\mathrm
E_{ij}\right)_{ab}=\delta_{ia}\delta_{jb}$ as
 \beq\label{vv705}
 \begin{array}{c}
  \displaystyle{
R^\hbar_{12}(z)=\sum\limits_{i,j,k.l=1}^N R^{\,\hbar}_{ij,kl}(z)
\,\mathrm E_{ij}\otimes \mathrm E_{kl}\,.
 }
 \end{array}
 \eq
Then it follows from (\ref{vv7026}) that
 \beq\label{vv706}
 \begin{array}{c}
  \displaystyle{
L^\eta(z)=L^\eta(z,S)=\sum\limits_{i,j,k,l=1}^N
R^{\,\eta}_{ij,kl}(z)\, \mathrm E_{ij}\, S_{lk}\,.
 }
 \end{array}
 \eq
The latter leads to the converse statement, i.e. having the
classical Lax matrix $L^\eta(z)$ we can find the quantum $R$-matrix
as
 \beq\label{vv707}
 \begin{array}{|c|}
  \hline\\
  \displaystyle{
 R^{\,\hbar}_{12}(z)=\frac{\p {L^\hbar_1}(z)}{\p
 S_{2}}=\sum\limits_{k,l=1}^N\frac{\p {L^\hbar}(z)}{\p
 S_{lk}}\otimes \mathrm E_{kl}
 }
  \\ \ \\
 \hline
  \end{array}
 \eq

{\bf The purpose of the paper} is twofold. The first one is to give
description of the relativistic classical tops arising from the
quantum $R$-matrices following (\ref{vv7026}). The second -- is to
derive new rational quantum $R$-matrix from the corresponding
relativistic top via (\ref{vv707}), which we obtain in a different
way - by applying gauge transformation of Hecke type
\cite{LOZ,LOSZ1,LOSZ4,SZ,AASZ} to the rational Ruijsenaars-Schneider
(RS) model \cite{Ruijs1}.


{\bf 1.  Relativistic classical tops from quantum $R$-matrices. }
Using local expansion of $L$-operator and $R$-matrix near $z=0$ we
get equations of motion related to the Hamiltonian $\mathcal
S_0=\tr(S)$ for the relativistic top in the form:
  \beq\label{vv708}
  \begin{array}{c}
  \displaystyle{
\p_{t_0}\,S=\{{\mathcal
S}_0\,,S\}\stackrel{(\ref{vv7026}-\ref{vv704})}{=}[S,J^\eta(S)]\,,
 }
 \end{array}
 \eq
where the inverse inertia tensor $J^\eta$ is the following linear
functional:
   \beq\label{vv709}
 \begin{array}{|c|}
  \hline\\
  \displaystyle{
 J^\eta(S)=\tr_2\left(\left(R_{12}^{\eta,(0)}-r_{12}^{(0)}\right)S_2\right)
   }
  \\ \ \\
 \hline
  \end{array}
  \eq
  with $R_{12}^{\eta,(0)}$ and $r_{12}^{(0)}$ be zero terms in the
  expansions (\ref{vv0011}), (\ref{vv0031}) near $z=0$ . See examples
  (\ref{vv3673}) and (\ref{vv71232}).
 These equations are presented in the Lax form
  \beq\label{vv710}
 \begin{array}{c}
  \displaystyle{
\p_{t_0}\,L^\eta (z)=\{{\mathcal S}_0\,,L^\eta
(z)\}\stackrel{(\ref{vv7026}-\ref{vv704})}{=}[ L^\eta(z)\,, M(z)]
 }
 \end{array}
 \eq
with $M$-operator defined in terms of the classical $r$-matrix:
 \beq\label{vv711}
 \begin{array}{c}
  \displaystyle{
M(z)=-\tr_2\left(r_{12}(z)S_2\right)\,.
 }
 \end{array}
 \eq
The latter $M$-operator appears to be equal (up to sign) to the
non-relativistic limit of the Lax matrix $L^\eta(z)$:
 \beq\label{vv7111}
 \begin{array}{c}
  \displaystyle{
L^\eta(z)= \eta^{-1}\,\frac{{\mathcal
 S}_0}{N}\,1_{N\times
 N}+l(z)+\eta\,{\mathcal M}(z)+O(\eta^2)\,,\ \ \
 l(z)=\tr_2\left(r_{12}(z)S_2\right)=-M(z)\,.
 }
 \end{array}
 \eq
Moreover, the next term in the expansion is the $M$-operator
 \beq\label{vv7112}
 \begin{array}{c}
  \displaystyle{
\p_t\, l(z,S)=[l(z,S),\mathcal M(z,S)]
 }
 \end{array}
 \eq
of the non-relativistic top given by equation
 \beq\label{vv71127}
 \begin{array}{c}
  \displaystyle{
\p_t\, S=[S,\mathrm J(S)]\,,\ \ \mathrm J(S)=\mathcal M(0,S)\,,
 }
 \end{array}
 \eq
where $\mathcal M(0)$ is the non-relativistic limit of $J^\eta(S)$
(\ref{vv709}).

The model (\ref{vv71127}) is bihamiltonian. It means that it can be
described by a pair of compatible Poisson structures. The first one
(the Poisson-Lie) is generated by the linear $r$-matrix structure
 \beq\label{vv7113}
 \begin{array}{c}
  \displaystyle{
 \{l_1(z)\,, l_2(w)\}=[\, l_1(z)+
l_2(w),r_{12}(z-w)]\,,
 }
 \end{array}
 \eq
and the second -- is by quadratic one
 \beq\label{vv7114}
 \begin{array}{c}
  \displaystyle{
 \{{\mathrm L}_1(z)\,, {\mathrm L}_2(w)\}=[\, {\mathrm L}_1(z)\,
{\mathrm L}_2(w),r_{12}(z-w)]
 }
 \end{array}
 \eq
with
 \beq\label{vv7115}
 \begin{array}{c}
  \displaystyle{
 {\mathrm L}(z,S)=\frac{{\mathrm s}_0}{N}\, 1 + l(z,S)-\frac{\tr\,
 l(z,S)}{N}\,1\,,
 }
 \end{array}
 \eq
where ${\mathrm s}_0$ is additional generator (of the classical
Sklyanin algebra). In the elliptic case (corresponding to the
Belavin-Drinfeld classical $r$-matrix \cite{BD}) this type of
bihamiltonian structure for (\ref{vv7114})-(\ref{vv7115}) was
described in \cite{KLO}.

Thus, we have two quadratic algebras -- (\ref{vv704}) with
$\eta$-dependent Lax operator (\ref{vv7026}), and (\ref{vv7114})
with $\eta$-independent Lax operator (\ref{vv7115}). Both algebras
are described by the same $r$-matrix. Then it is natural to expect a
relation between $L^\eta(z,S)$ and $\mathrm L(z,S)$. It can be
written explicitly:
 \beq\label{vv7116}
 \begin{array}{|c|}
  \hline\\
  \displaystyle{
 L^\eta\Big(z+\eta_0,{\mathrm
L}(-\eta_0,S)\Big)=\frac{\tr L^\eta\left(z+\eta_0,S\right)}{\tr
S}\,{\mathrm L}(z,S)
 }
  \\ \ \\
 \hline
  \end{array}
 \eq
 where ${\mathrm
L}(-\eta_0,S)$ is inserted into $L^\eta(z,S)$ as the second
argument, $\eta_0=\eta_0(\eta)$ is a zero of function $\tr
L^\eta\left(z,S\right)/\tr
 S$. In notations of this paper $\eta_0=-\eta$ in the elliptic case and
 $\eta_0=-\eta/N$ in the rational one.
Then the change of variables from the $\eta$-independent description
(\ref{vv7115}) to the
 $\eta$-dependent description (\ref{vv7026}) can
 be written as
 \beq\label{vv71167}
 \begin{array}{c}
  \displaystyle{
 S\ \to\ -({\eta_0}/{\eta})\,\mathrm L(-\eta_0,S)\,.
 }
  \end{array}
 \eq
Relation (\ref{vv7116}) allows also to find the $M$-operator for
(\ref{vv7115}) as
 \beq\label{vv71162}
 \begin{array}{c}
  \displaystyle{
 \mathrm M(z,S)=-({\eta_0}/{\eta})\, J^{-(\eta/\eta_0)z}(\mathrm  L(z,S))\,.
 }
  \end{array}
 \eq
%

The simplest example of the relativistic top is obtained in the
elliptic case,
 where the quantum $R$-matrix is the Belavin's one \cite{Belavin}.
 In ${\rm gl}_2$ case it coincides with the Baxter's one. Then for
 $S=\sum\limits_{a=0}^3  S_a \si_a$, where $\si_a$ are the Pauli
 matrices ($\si_0=1$)
 \beq\label{vv71232}
 \begin{array}{c}
  \displaystyle{
J^\eta(S)=\sum\limits_{a=0}^3 J^\eta_a\, S_a\,\si_a\,,\ \
J^\eta_0=E_1(\eta)\,,\ \
J^\eta_\al=E_1(\eta+\om_\al)-E_1(\om_\al)\,,\ \ \al=1\,,2\,,3\,,
 }
 \\
   \displaystyle{
E_1(z)=\p_z\log\vth(z|\tau)\,,\ \ \om_1=\tau/2\,,\
\om_2=(1+\tau)/2\,,\ \om_3=1/2\,.
 }
 \end{array}
 \eq
 In the elliptic
case
 we also consider the large $N$ limit to the elliptic
 hydrodynamics \cite{KLO,O} by passage to the noncommutative torus
 description:
 \beq\label{vv7118}
   \begin{array}{|c|}
  \hline\\
 \displaystyle{
\p_t S(x)=\ad^*_{J^\eta(S)(x)}S(x)=[S(x),J^\eta(S)(x)]_\te
  }
   \\ \ \\
 \hline
 \end{array}
 \eq
 where
 $
 [f(x),g(x)]_\te=\te^{-1} (f\star g- g\star f)
 $ is the Moyal bracket ($\star$ is the Moyal product) and $J^\eta(S)$ is the pseudo-differential operator given in
 (\ref{J}) (cf. (\ref{vv413})).

{\bf 2. Quantum rational $R$-matrix.} We propose the factorized form
for the rational Ruijsenaars-Schneider (RS) Lax matrix
 \beq\label{vv712}
 \begin{array}{c}
  \displaystyle{
{L}^{\hbox{\tiny{RS}}}(z)=g^{-1}(z)\,g(z+\eta)\,e^{P/c}\,,
 }
 \end{array}
 \eq
where $c$ is the light speed, $P$ is a diagonal matrix of the RS
particles momenta, and $g(z)$ is the matrix depending on the RS
particles coordinates $q$. The latter was introduced in \cite{AASZ},
where the non-relativistic rational top was constructed similarly
starting from the rational Calogero-Moser (CM) model
\cite{Calogero}. The transformation $g(z)$ is known for the quantum
elliptic and trigonometric RS models \cite{Hasegawa12,Zabrodin1}
(where the quantum IRF-Vertex correspondence was described). The
rational one was mentioned in \cite{AASZ}\footnote{These gauge
transformations underly the Symplectic Hecke Correspondence
\cite{LOZ} (see also \cite{LOSZ1,SZ}) for the classical integrable
systems in the Hitchin approach.}. By performing the gauge
transformation
 \beq\label{vv713}
 \begin{array}{c}
  \displaystyle{
L^\eta(z)=g(z){
L}^{\hbox{\tiny{RS}}}(z)g^{-1}(z)=g(z+\eta)\,e^{P/c}\,g^{-1}(z)
 }
 \end{array}
 \eq
and re-expressing $L^\eta(z)$ in terms of its residue we come to the
relativistic rational top. It corresponds to some special values of
the Casimir functions, while arbitrary values are related in the
same way to the spin RS model \cite{KrichZabr} (see also
\cite{Arut}). The answer is given in Section \ref{lax}.

Then using (\ref{vv7026}) we obtain rational unitary quantum
R-matrix. In ${\rm gl}_2$ case  it is the 11-vertex $R$-matrix
 \beq\label{vv714}
 \begin{array}{c}
  \displaystyle{R^\hbar(z)=
 \left( \begin{array}{cccc} {\hbar}^{-1}+{z}^{-1}&0&0&0
\\\noalign{\medskip}-\hbar-z&{\hbar}^{-1}&{z}^{-1}&0\\\noalign{\medskip}
-\hbar-z&{z}^{-1}&{\hbar}^{-1}&0\\\noalign{\medskip}-{\hbar}^{3}-2\,z{\hbar}^{2}-2\,\hbar
\,{z}^{2}-{z}^{3}&\hbar+z&\hbar+z&{\hbar}^{-1}+{z}^{-1}
\end{array} \right)
  }
 \end{array}
 \eq
obtained previously in \cite{Cherednik}\footnote{It can be also
obtained \cite{Smirnov} by applying special limiting procedure to
the Baxter elliptic $R$-matrix \cite{Baxter}.}. In Section
\ref{quant} we obtain ${\rm gl}_N$ generalization of
(\ref{vv714}). 
Introduction of $\epsilon$ parameter as
 \beq\label{vv7140}
 \begin{array}{c}
  \displaystyle{R^{\hbar,\epsilon}(z)=\epsilon\,R^{\,\epsilon\hbar}(\epsilon z)
  }
 \end{array}
 \eq
 allows to interpret it as deformation of the XXX $R$-matrix
 \beq\label{vv71400}
 \begin{array}{c}
  \displaystyle{
  \lim\limits_{\epsilon\rightarrow 0}R^{\hbar,\epsilon}(z)=R^{\hbox{\tiny{XXX}}}(z)=\frac{1}{\hbar}\,
  1\otimes 1+\frac{1}{z}\,P_{12}\,,\ \ P_{12}=\sum\limits_{i,j=1}^N\, \mathrm
  E_{ij}\otimes \mathrm E_{ji}\,.
  }
 \end{array}
 \eq
Notice that in this limit the relativistic top  (\ref{vv708}),
(\ref{vv709}) becomes free mechanical system
 in the sense that
$L^\eta(z)=\eta^{-1}\mathcal S_0\,1+z^{-1}S$, and equations of
motion are trivial $\dot S=0$. Therefore, the parameter $\epsilon$
can be also treated as an alternative definition of the coupling
constant.

The Lax matrix (\ref{vv713}) (which is gauge equivalent to the RS
model) emerge from explicit change of variables:
 \beq\label{vv7142}
 \begin{array}{c}
  \displaystyle{
  { L^\eta}(z)=\tr_2 \left(R^{\,\eta}_{12}(z) S_2\right)\,,
  }
  \\ \ \\
  \displaystyle{
  S_{ij}(\bfq,{\bf p})=\sum_{m=1}^{N}\,\frac{({
q}_{m}+\eta)^{\,\varrho(i)} e^{p_{m}/c} }{ \prod\limits_{k\neq
m}^{\,} ({ q}_{m}-{
q}_{k})}\,\,\,(-1)^{\varrho(j)}\,\sigma_{\varrho(j)}(\bfq)\,,
  }
 \end{array}
 \eq
where $\varrho(i)=(i-1)\delta_{i\leq N-1}+N\delta_{iN}$ (see
(\ref{vv922})), while $\sigma_j(q)$ are elementary symmetric
functions (\ref{vv510})-(\ref{vv512}). The case (\ref{vv7142})
corresponds to rank one matrix $S$ and to special values of the
Casimir function $\det L^\eta(z)$ of Poisson brackets (\ref{vv704}).
In the (quantum) elliptic and trigonometric cases the
(\ref{vv7142})-type formulae for $\hat S=\hat
S(\bfq,\frac{\p}{\p\bfq})$ can be found in
\cite{Sklyanin,Hasegawa12,Zabrodin1} (see also \cite{KrichZabr}).

In general case $L^\eta(z)=\tr_2 \left(R^{\,\eta}_{12}(z)
S_2\right)$, where all $S_{ij}$ are independent variables. For
non-relativistic models it was shown in \cite{LOZ} that the top
models on the special coadjoint orbit are gauge equivalent to
Calogero-Moser (CM) systems \cite{Calogero} while generic orbits
correspond to their spin generalizations. In the same way, the
generic relativistic top can be treated as alternative form of the
spin RS model \cite{KrichZabr}. The gauge transformations used in
(\ref{vv713}) are of the same form as in non-relativistic case,
where they play the role of modifications of the underlying Higgs
bundles. Hence, we deal with the relativistic version of the
Symplectic Hecke Correspondence. It allows us to obtain the
non-dynamical quantum $R$-matrix instead of direct usage of the
quantum IRF-Vertex Correspondence \cite{irf-vertex}. In this
respect, we realize the latter correspondence by means of the
relativistic version of the classical (Symplectic Hecke) one. It is
also interesting to mention that in view of (\ref{vv7116}),
(\ref{vv71167}) we obtain the same form of equations (\ref{vv71127})
for the (spin) RS and (spin) CM models


{\bf Spin chains and Gaudin models} related to the 11-vertex
rational $R$-matrix (\ref{vv714}) and its classical limit are
obtained straightforwardly. As an example we get the ${\rm gl}_2$
Gaudin model Hamiltonians emerging from non-relativistic limit of
the inhomogeneous chain:
 \beq\label{vv4487}
 \begin{array}{c}
  \displaystyle{
  h_a=\sum\limits_{c\neq a}^n\,\tr \left( \hat S^a\,l(z_a-z_c,
 \hat S^c)\right)=\sum\limits_{c\neq a}^n\,\tr_{12}\left(r_{12}(z_a-z_c) \hat S^a_1  \hat S^c_2\right)=
 }
 \end{array}
 \eq
$$
\sum\limits_{c\neq a}^n\,\frac{\tr(\hat S^a \hat S^c)}{z_a-z_c}
  -(z_a-z_c)\left(\hat S_{12}^a(\hat S_{11}^c-\hat S_{22}^c)+\hat S_{12}^c(\hat S_{11}^a-\hat S_{22}^a)\right)
  -(z_a-z_c)^3\,\hat S_{12}^a\hat S_{12}^c\,.
 $$
 Notice that the first term corresponds to the standard rational (XXX) Gaudin
 Hamiltonians.

The 11-vertex model is defined by the $R$-matrix (\ref{vv714}). The
quantum local Hamiltonian of the homogeneous periodic spin (1/2)
chain on $n$ sites is of the form:
 \beq\label{vv4497}
 \begin{array}{c}
  \displaystyle{
  H^{\hbox{\tiny{local}}}=\sum\limits_{k=1}^n\, P_{k,k+1}-\eta^2 \mathrm E_{21}^{k}\otimes(\mathrm E_{11}^{k+1}-\mathrm E_{22}^{k+1})
  -\eta^2 (\mathrm E_{11}^{k}-\mathrm E_{22}^{k})\otimes\mathrm
  E_{21}^{k+1}-\eta^4 \mathrm E_{21}^{k}\otimes\mathrm
  E_{21}^{k+1}\,,
 }
 \end{array}
 \eq
 where $\mathrm E_{ij}^{n+1}=\mathrm E_{ij}^1$ (here we use the dual generators $\mathrm
 E_{ij}$: $(\mathrm E_{ij})_{ab}=\delta_{ia}\delta_{jb}$, $S_{ji}=\tr (\mathrm E_{ij}\hat
 S)$).
It is a deformation of the XXX spin chain (see also \cite{Khor},
where this type of deformation was obtained using different
$R$-matrix)
 described by the first term in (\ref{vv4497}):
 \beq\label{vv4498}
 \begin{array}{c}
  \displaystyle{
  H^{\hbox{\tiny{XXX}}}=\sum\limits_{k=1}^n\, P_{k,k+1}\,,\ \ \
  P_{k,k+1}=\mathrm E_{11}^{k}\otimes\mathrm
  E_{11}^{k+1}+\mathrm E_{12}^{k}\otimes\mathrm
  E_{21}^{k+1}+\mathrm E_{21}^{k}\otimes\mathrm
  E_{12}^{k+1}+\mathrm E_{22}^{k}\otimes\mathrm
  E_{22}^{k+1}\,.
 }
 \end{array}
 \eq

\vskip3mm

{\small

\noindent {\bf Acknowledgments.} The work was supported by RFBR
grants 12-02-00594 (A.L. and M.O.) and 12-01-00482 (A.Z.). The work
of A.L. was partially supported by AG Laboratory GU-HSE, RF
government grant, ag. 11 11.G34.31.0023. The work of A.Z. was
partially supported by the D. Zimin's fund "Dynasty" and by the
Program of RAS "Basic Problems of the Nonlinear Dynamics in
Mathematical and Physical Sciences"  $\Pi$19.

}




\section{Sklyanin algebras and classical integrable systems}
\setcounter{equation}{0}


\subsection{Quantum Sklyanin algebra}

Let  the quantum $L$-operator (\ref{vv001}) has the following
expansions in spectral parameter near $z=0$:
 \beq\label{vv0021}
 \begin{array}{c}
  \displaystyle{
 {\hat L^\eta}(z)=L^\eta(z,\hat S)=\sum\limits_{k=-1}^\infty z^k\,{ L^{\eta,(k)}}(\hat S)=\frac{1}{z} {\hat S}+{ L^{\eta,(0)}}(\hat S)+z\,{
 L^{\eta,(1)}}(\hat S)+O(z^2)\,,
 }
 \end{array}
 \eq
where the residue ${\hat S}$ is ${\rm gl}_N$-valued operator and
 $L^{\eta,(k)}$ are linear functionals of $\hat S$, i.e.
 \beq\label{vv0022}
 \begin{array}{c}
  \displaystyle{
 {\hat L^\eta}(z)=\sum\limits_{a,b}\sum\limits_{k=-1}^\infty z^k \stackrel{k}{\mathcal
 R^\eta}_{a,b}\,T_a\,{\hat S}_b\,,\ \ \  {L^{\eta,(k)}}(\hat S)=\sum\limits_{a,b}\, \stackrel{k}{\mathcal
 R^\eta}_{a,b}\,T_a\,{\hat S}_b
 }
 \end{array}
 \eq
 in  some  basis $\{T_a\}$ of ${\rm gl}_N$. The coefficients $\stackrel{k}{\mathcal
 R^\eta}_{a,b}$ are functions of a free constant parameter $\eta$ which role is explained below. Due to (\ref{vv001})
 the matrix elements ${\hat
 S}_b$ satisfy quadratic relations of ${\mathcal A}_{\hbar,\eta}^{\hbox{\tiny{Skl}}}$ (see (\ref{vv007})) such as Sklyanin
 algebra \cite{Sklyanin}
 or its different extensions \cite{OF,Rosen,CLOZ,Ruijs2}.
Notice that the representation space of
 operators $\hat S_a$ is not fixed yet.

  Similarly to (\ref{vv0021}) and (\ref{vv0022}) let the $R$-matrix
  be of the form:
 \beq\label{vv0011}
 \begin{array}{c}
  \displaystyle{
 R^\hbar_{12}(z)=\sum\limits_{k=-1}^\infty z^k\,R^{\hbar,(k)}_{12}=\frac{1}{z}\,P_{12}+R^{\hbar,(0)}_{12}+z\,R^{\hbar,(1)}_{12}+O(z^2)\,,
 \ \ \ R^{\hbar,(k)}_{12}\in {{\rm gl}_N}^{\otimes 2}\,,
 }
 \end{array}
 \eq
 \beq\label{vv0024}
 \begin{array}{c}
  \displaystyle{
 R^{\hbar,(k)}_{12}=\sum\limits_{a,b} \stackrel{k}{\mathcal
 R^\hbar}_{a,b}\,T_a\otimes T_{-b}\,,
 }
 \end{array}
 \eq
where the generators $T_{-b}$ are dual to $T_b$: $\tr\left(T_a
T_b\right)=\delta_{a+b}$, and
$R^{\hbar,(-1)}_{12}=P_{12}=\sum\limits_a T_a\otimes T_{-a}$ is the
permutation operator. Formulae (\ref{vv0011}) and (\ref{vv0024})
imply the following simple link
 between $L$-operator and $R$-matrix:
 \beq\label{vv0025}
  \begin{array}{c}
  \displaystyle{
 {\hat L^\eta}(z)=\tr_2 \left(R^{\,\eta}_{12}(z)\hat S_2\right)\,.
 }
 \end{array}
 \eq
%
It is important to mention that we deal with two constants $\hbar$
and $\eta$ (\ref{vv001}). While $\hbar$ plays the role of the Planck
constant, the  parameter $\eta$ will be shown to describe
relativistic deformation in the sense of Ruijsenaars.
%
%

Using notations of (\ref{vv0021}) and (\ref{vv0011})  it easy to
write down the quadratic relations of ${\mathcal
A}_{\hbar,\eta}^{\hbox{\tiny{Skl}}}$. Indeed, consider residue of
(\ref{vv002}) at $w=0$:
 \beq\label{vv0072}
 \begin{array}{c}
  \displaystyle{
R^\hbar_{12}(z)\,{\hat L^\eta}_1(z)\,{\hat S}_2={\hat S}_2\,{\hat
 L^\eta}_1(z)\,R^\hbar_{12}(z)\,.
 }
 \end{array}
 \eq
Expanding this equation near $z=0$ we get identity $P_{12}{\hat
S}_1{\hat S}_2={\hat S}_2{\hat S}_1 P_{12}$ for $z^{-2}$ terms while
the coefficients behind $z^{-1}$ give rise to the Sklyanin
algebra\footnote{In his original paper \cite{Sklyanin} Sklyanin used
 $\eta=\hbar$.}:
 \beq\label{vv007}
 \begin{array}{c}
  \displaystyle{
{\mathcal A}_{\hbar,\eta}^{\hbox{\tiny{Skl}}}:\ \ \
P_{12}\,L^{\eta,(0)}(\hat S)_1\, {\hat S}_2+R^{\hbar,(0)}_{12}\,
{\hat S}_1\, {\hat S}_2={\hat S}_2\,L^{\eta,(0)}(\hat S)_1\, P_{12}+
{\hat S}_2\, {\hat S}_1\, R^{\hbar,(0)}_{12}\,.
 }
 \end{array}
 \eq
 A typical representative for
(\ref{vv0011})-type of solutions of the Yang-Baxter equation
(\ref{vv002}) is the Belavin's elliptic $R$-matrix. It is considered
in Section \ref{ell}.

 It follows from (\ref{vv0024}) that the Sklyanin algebra
${\mathcal A}_{\hbar,\hbar}^{\hbox{\tiny{Skl}}}$ (with $\eta=\hbar$)
has finite-dimensional representation\footnote{In ${\rm gl}_2$ case
the original Sklyanin algebra has simple representation in terms of
the Pauli matrices $\hat S_a=\sigma_{-a}=\sigma_{a}$, $a=0,1,2,3$,
with $\sigma_0=1_{2\times 2}$, which are used as basis $\{T_a\}$.}
 \beq\label{vv0023}
 \begin{array}{c}
  \displaystyle{
\rho\left({\mathcal A}_{\hbar,\hbar}^{\hbox{\tiny{Skl}}}\right):\ \
\ \rho({\hat S}_a)=T_{-a}\in{\rm gl}_N\,,
 }
 \end{array}
 \eq

\noindent Then
 \beq\label{vv00231}
 \begin{array}{c}
  \displaystyle{
 R^{\hbar,(k)}_{12}=\rho\left({L^{\hbar,(k)}}(\hat S)\right)\,.
 }
 \end{array}
 \eq
%
With this definition the quantum Yang-Baxter equation  (\ref{vv002})
coincides with exchange relations (\ref{vv001}) in representation
(\ref{vv0023}).


\subsection{Classical limit}\label{CLalim}

\noindent {\bf Quantum $R$-matrix.} In the classical limit
$\hbar\rightarrow 0$ the operators $\hat S_a$ become $\mathbb
C$-valued coordinates on the phase space of an integrable system
described by the Lax matrix $L^\eta(z,S)$. Notice that relation
(\ref{vv0025}) remains intact at classical level, i.e.
 \beq\label{vv0027}
  \begin{array}{c}
  \displaystyle{
 {L^\eta}(z,S)=\tr_2 \left(R^{\,\eta}_{12}(z) S_2\right)=\sum\limits_{a,b}\sum\limits_{k=-1}^\infty z^k \stackrel{k}{\mathcal
 R^\eta}_{a,b}\,T_a\,{S}_b\,.
 }
 \end{array}
 \eq
Therefore, having the classical Lax matrix $L^{\eta}(z)$ of the
described type we can compute the quantum $R$-matrix in the
following way:
 \beq\label{vv0028}
  \begin{array}{c}
  \displaystyle{
 R^{\,\hbar}_{12}(z)=\sum\limits_{b}\frac{\p {L^\hbar}(z,S)}{\p
 S_b}\otimes T_{-b}
 }
 \end{array}
 \eq
 We will use this formula in Section \ref{rat} for derivation of the
 rational $R$-matrix.

\vskip3mm

\noindent {\bf Classical $r$-matrix.} Let the quantum $R$-matrix has
the following expansion in the Planck constant $\hbar$:
 \beq\label{vv003}
 \begin{array}{c}
  \displaystyle{
 R^\hbar_{12}(z)=\frac{1}{\hbar}\,1\otimes 1+r_{12}(z)+\hbar\,
 r'_{12}(z)+O(\hbar^2)\in {{\rm gl}_N}^{\otimes 2}\,,
 }
 \end{array}
 \eq
 where 
  \beq\label{vv0031}
 \begin{array}{c}
  \displaystyle{
 r_{12}(z)=\frac{P_{12}}{z}+r_{12}^{(0)}+O(z)\,,\ \ \
 r'_{12}(z)=r'_{12}(0)+O(z)\,.
 }
 \end{array}
 \eq
and $r_{12}^{(0)}$ comes from $R^{\hbar,(0)}_{12}$ in
(\ref{vv0011}):
  \beq\label{vv0032}
 \begin{array}{c}
  \displaystyle{
 R^{\hbar,(0)}_{12}=\frac{1}{\hbar}\,1\otimes
1+r_{12}^{(0)}+O(\hbar)\,.
 }
 \end{array}
 \eq
 The term $r_{12}(z)$ is the classical $r$-matrix. It
is skew-symmetric
 \beq\label{vv0033}
 \begin{array}{c}
  \displaystyle{
  r_{12}(z)=-r_{21}(-z)\,,
 }
 \end{array}
 \eq
 \beq\label{vv0034}
 \begin{array}{c}
  \displaystyle{
\stackrel{k}{\mathcal
 R^0}_{a,b} =(-1)^{k+1} \stackrel{k}{\mathcal
 R^0}_{b,a}\,,\ \ \ \ \ \stackrel{0}{\mathcal
 R^0}=\lim\limits_{\hbar\rightarrow 0}(\stackrel{0}{\mathcal
 R^\hbar}-\hbar^{-1}\,1\otimes 1)
 }
 \end{array}
 \eq
and satisfies the classical Yang-Baxter equation:
 \beq\label{vv0041}
 \begin{array}{c}
  \displaystyle{
[r_{12}(z-w),r_{13}(z)]+[r_{12}(z-w),r_{23}(w)]+[r_{13}(z),r_{23}(w)]=0\,.
 }
 \end{array}
 \eq
The latter arises from (\ref{vv002}) and (\ref{vv003}).
Similarly, by substituting (\ref{vv003}) into (\ref{vv001}) we come
to quadratic Poisson structure
 \beq\label{vv004}
 \begin{array}{c}
  \displaystyle{
\lim\limits_{\hbar\rightarrow 0} \,\frac{1}{\hbar}\,[{\hat
L^\eta}_1(z)\,,{\hat L^\eta}_2(w)]:=\{L^\eta_1(z)\,, L^\eta_2(w)\}=[
L^\eta_1(z)\, L^\eta_2(w),r_{12}(z-w)]\,,
 }
 \end{array}
 \eq
 where $L^\eta(z)$ is the classical $L$-operator (the Lax matrix).



\subsection{Relativistic top}

Let us define the {\em relativistic top} as an integrable model
described by the Lax matrix (\ref{vv0027}) and the $r$-matrix
structure (\ref{vv004}):
 \beq\label{vv0084}
 \begin{array}{c}
  \displaystyle{
\{L^\eta_1(z)\,, L^\eta_2(w)\}=[ L^\eta_1(z)\,
L^\eta_2(w),r_{12}(z-w)]\,,
 }
 \\ \ \\
  \displaystyle{
{L^\eta}(z)=\tr_2 \left(R^{\,\eta}_{12}(z) S_2\right)=\frac{1}{z}
{S}+{ L^{\eta,(0)}}(S)+z\,{
 L^{\eta,(1)}}(S)+O(z^2)\,.
 }
 \end{array}
 \eq
We will see that equations of motion have the form of the integrable
multidimensional Euler (or Euler-Arnold) top. On the other hand, it
will be shown below that the parameter $\eta$ plays the same role as
the relativistic deformation parameter in the Ruijsenaars-Schneider
generalization of Calogero-Moser models. This is why we call these
type of models relativistic tops\footnote{Our approach is in
agreement with the one considered in \cite{Hanson} for relativistic
particles and rotators (in an external fields and on curved spaces),
where the authors also used the term relativistic top. Our case
corresponds to their free top.}.

\noindent{\bf Classical Sklyanin algebra.} The phase space is
parameterized by $N^2$ coordinates $\{S_a\}$. It is equipped with
the following quadratic Poisson structure:
 \beq\label{vv009}
 \begin{array}{c}
  \displaystyle{
{\mathcal A}_{\hbar=0,\eta}^{\hbox{\tiny{Skl}}}:\ \ \
\{S_1,S_2\}=[S_1S_2,r_{12}^{(0)}]+[L^{\eta,(0)}(S)_1\,S_2,P_{12}]\,,
 }
 \end{array}
 \eq
where $r_{12}^{(0)}$ is defined in (\ref{vv0031}) and
$L^{\eta,(0)}(S)$  in (\ref{vv0021}) and (\ref{vv0084}). The
brackets (\ref{vv009}) can be obtained both -- from the quantum
algebra (\ref{vv007}) (by taking the classical limit (\ref{vv003}))
or from (\ref{vv0084}) by computing residue at $w=0$
 \beq\label{vv010}
 \begin{array}{c}
  \displaystyle{
\{L^\eta_1(z)\,, S_2\}=[ L^\eta_1(z)\, S_2,r_{12}(z)]
 }
 \end{array}
 \eq
and evaluating the coefficient in front of $z^{-1}$. The Poisson
brackets are degenerated. In order to restrict it on a symplectic
leaf we need to fix Casimir functions $C_k(S)$. They appear as
coefficients in the expansion of $\det L^\eta(z)$ which is known to
be central element for the Poisson brackets (\ref{vv004}):
 \beq\label{vv011}
 \begin{array}{c}
  \displaystyle{
\det L^\eta(z)=\sum\limits_{k=-N}^\infty z^k C_k(S)\,.
 }
 \end{array}
 \eq
The number of independent Casimir functions (in general) equals $N$.
They can be accumulated from coefficients in front of nonpositive
powers of $z$ in (\ref{vv011}) (others are dependent). The
Hamiltonians (including the Casimir functions) can be computed from
the expansion near $z=0$ of
 \beq\label{vv012}
  \begin{array}{c}
  \displaystyle{
\frac{1}{k}\tr\left(L^\eta(z)\right)^k=\frac{1}{k}\tr_{0,1,\ldots,k}\left(R^\eta_{01}(z)\ldots
R^\eta_{0k}(z)\, S_1\ldots S_k\right)=
 }
 \\ \ \\
  \displaystyle{=\frac{1}{z^k}H_{k,k}+\frac{1}{z^{k-1}}H_{k,k-1}+...+
 H_{k,0}+...\,,\
\ \ k=1\,...\,N
 }
 \end{array}
 \eq
or from the spectral curve
 \beq\label{vv013}
  \begin{array}{c}
  \displaystyle{
\det\left(\lambda-L^\eta(z)\right)=0\,.
 }
 \end{array}
 \eq
The number of Casimir functions $N$ should be subtracted from the
number of independent Hamiltonians (\ref{vv012}) $N(N+1)/2$. This
gives $N(N-1)/2$ for the Hamiltonians only. It equals to the half of
dimension of a general symplectic leaf. The Poisson commutativity of
the Hamiltonians $H_{k,l}$ is guaranteed by (\ref{vv0084}).
Therefore, the model is integrable in the Liouville-Arnold sense.


%

\noindent{\bf Equations of motion and Lax pair.} The simplest
Hamiltonian is given by
 \beq\label{vv014}
  \begin{array}{c}
  \displaystyle{
{\mathcal S}_0:=H_{1,1}=\tr (S) \,.
 }
 \end{array}
 \eq
 To get equations of motion let us compute the trace over the second
 component (in tensor product ${{\rm gl}_N}^{\otimes 2}$) of
 (\ref{vv009}). It leads to the top-like equations
  \beq\label{vv0142}
  \begin{array}{c}
  \displaystyle{
\p_{t_0}\,S=\{{\mathcal S}_0\,,S\}=[S,J^\eta(S)]\,,
 }
 \end{array}
 \eq
where the inverse inertia tensor $J^\eta$ is the following linear
functional of $S$:
   \beq\label{vv0143}
  \begin{array}{c}
  \displaystyle{
 J^\eta:\ \ S\ \rightarrow\ J^\eta(S)_1=-\tr_2(r_{12}^{(0)}\,S_2)+L^{\eta,(0)}(S)_1
 \stackrel{(\ref{vv0027})}{=}\tr_2\left(\left(R_{12}^{\eta,(0)}-r_{12}^{(0)}\right)S_2\right)\,.
   }
   \end{array}
  \eq
In a similar way, by applying $\tr_2$ to the both parts of
(\ref{vv010}) we get equations of motion (\ref{vv0142}) in the Lax
form:
 \beq\label{vv015}
 \begin{array}{c}
  \displaystyle{
\p_{t_0}\,L^\eta (z)=\{{\mathcal S}_0\,,L^\eta (z)\}=[ L^\eta(z)\,,
M(z)]\,,
 }
 \end{array}
 \eq
where the $M$-operator equals
 \beq\label{vv016}
 \begin{array}{c}
  \displaystyle{
M_1(z)=-\tr_2\left(r_{12}(z)S_2\right)\,.
 }
 \end{array}
 \eq
Notice that it is independent of $\eta$. As we will see below the
$M$-operator in this description coincides  with non-relativistic
Lax matrix.

\subsection{Non-relativistic limit}\label{lin}


The non-relativistic limit $\eta\rightarrow 0$ is similar to the
classical one due  (\ref{vv0027}). It follows from (\ref{vv003}) and
(\ref{vv0027}) that
 \beq\label{vv017}
 \begin{array}{c}
  \displaystyle{
 L^\eta(z)= \frac{{\mathcal
 S}_0}{N\eta}\,1_{N\times
 N}+\tr_2\left(r_{12}(z)S_2\right)+\eta\,\tr_2\left(r'_{12}(z)S_2\right)+O(\eta^2)\,,
 }
 \end{array}
 \eq
 \beq\label{vv018}
 \begin{array}{c}
  \displaystyle{
 L^{\eta,(0)}= \frac{{\mathcal
 S}_0}{N\eta}\,1_{N\times N}+\tr_2\left(r_{12}^{(0)}S_2\right)+\eta\,\tr_2\left(r'_{12}(0)S_2\right)+O(\eta^2)\,.
 }
 \end{array}
 \eq
Plugging (\ref{vv017}) and (\ref{vv018})  into (\ref{vv0084}) and
(\ref{vv009}) respectively we get
 \beq\label{vv019}
 \begin{array}{c}
  \displaystyle{
 \{L^\eta_1(z)\,, L^\eta_2(w)\}=\frac{{\mathcal
 S}_0}{N\eta}\,[\, l_1(z)+
l_2(w),r_{12}(z-w)]+[\, l_1(z)\, l_2(w),r_{12}(z-w)]+O(\eta)\,,
 }
 \end{array}
 \eq
where
 \beq\label{vv020}
 \begin{array}{c}
  \displaystyle{
 l_1(z):=\tr_2\left(r_{12}(z)S_2\right)\,,
 }
 \end{array}
 \eq
  \beq\label{vv0201}
 \begin{array}{c}
   \displaystyle{
 l(z)=\frac{1}{z}S+l^{(0)}(S)+z\,l^{(1)}(S)+O(z)\,,\ \
 \ l^{(0)}(S)_1=\tr_2\left(r_{12}^{(0)}S_2\right)\,.
 }
 \end{array}
 \eq
and
 \beq\label{vv021}
 \begin{array}{c}
  \displaystyle{
\{S_1,S_2\}=\frac{{\mathcal
 S}_0}{N\eta}\,[S_2,P_{12}] +[S_1S_2,r_{12}^{(0)}]+[\tr_3
 (r_{13}^{(0)}S_3)\,S_2,P_{12}] + O(\eta)\,,
 }
 \end{array}
 \eq
where $\tr_3
 (r_{13}^{(0)}S_3)$ is just $ l^{(0)}(S)_1$ as in (\ref{vv0201}).

When $\eta\rightarrow 0$ the leading term in (\ref{vv021}) is the
linear Poisson-Lie structure on ${{\rm gl}}_N^*$ Lie coalgebra. The
generator ${\mathcal S}_0=\tr S$ is the Casimir function of the
latter brackets. Let us fix it as ${\mathcal S}_0=N$ and set
 \beq\label{vv0224}
 \begin{array}{c}
  \displaystyle{
\{S_1,S_2\}_{Lie}:=\lim\limits_{\eta\rightarrow
0}\eta\,\{S_1,S_2\}\,.
 }
 \end{array}
 \eq
 Then
 \beq\label{vv022}
 \begin{array}{c}
  \displaystyle{
\{S_1,S_2\}_{Lie}=[S_2,P_{12}] \,.
 }
 \end{array}
 \eq
 In the same way the linear $r$-matrix structure is obtained at the level of Lax matrices:
 \beq\label{vv023}
 \begin{array}{c}
  \displaystyle{
\{l_1(z)\,, l_2(w)\}_{Lie}=[\, l_1(z)+ l_2(w),r_{12}(z-w)]\,.
 }
 \end{array}
 \eq
\noindent {\bf Non-relativistic top.} We will refer to an integrable
model described by the Lax matrix (\ref{vv020}) and the Poisson
structure (\ref{vv022}), (\ref{vv023}) as the {\em  non-relativistic
top}. The phase space is the coadjoint orbit of ${{\rm GL}}_N$ Lie
group. It is equipped with the linear Poisson-Lie structure on
${{\rm gl}}_N^*$. For example, using the standard basis of matrices
$\left({\mathrm E}_{ij}\right)_{ab}=\delta_{ia}\delta_{jb}$
(\ref{vv022}) acquires the from
$\{S_{ij},S_{kl}\}=\delta_{il}S_{kj}-\delta_{kj}S_{il}$. A general
symplectic leaf is obtained by fixation of eigenvalues of $S$ or the
Casimir functions $c_k=\frac{1}{k}\tr(S^k)$, $k=1...N$. The
Hamiltonians appear as in (\ref{vv012}):
 \beq\label{vv024}
  \begin{array}{c}
  \displaystyle{
\frac{1}{k}\tr\left(l(z)\right)^k=\frac{1}{k}\tr_{0,1,\ldots,k}\left(r_{01}(z)\ldots
r_{0k}(z)\, S_1\ldots S_k\right)=
 }
 \\ \ \\
  \displaystyle{
=\frac{1}{z^k}h_{k,k}+\frac{1}{z^{k-1}}h_{k,k-1}+...+
 h_{k,0}+...\,,\
\ \ k=1\,...\,N\,.
 }
 \end{array}
 \eq
It is easy to see that $h_{k,k}=c_k$. The Poisson commutativity of
the Hamiltonians $h_{k,l}$ is guaranteed by (\ref{vv023}). The
$M$-operators corresponding to the Hamiltonians $h_{k,l}$ are
evaluated by expansion of $-\tr_2(r_{12}(z-w)l_2^{k-1}(w))$ (see
 \cite{Babel}). An alternative way is given in
 (\ref{vv039})-(\ref{vv0393}).

Notice that the Poisson brackets (\ref{vv022}) follows from
(\ref{vv023}) and local expansion from (\ref{vv0201}). To get
(\ref{vv022}) one should substitute expansion (\ref{vv0201}) into
(\ref{vv023})  and compute the residue at $w=0$ and then at $z=0$.

Similarly to (\ref{vv0027}), we have a simple link between the Lax
matrix and classical $r$-matrix given by (\ref{vv020}). Substitution
of (\ref{vv020}) into (\ref{vv023}) gives rise to the classical
Yang-Baxter equation (\ref{vv0041}) (it follows from the Jacobi
identity for the Poisson brackets (\ref{vv023}) as well).
    By analogy with  (\ref{vv0028}) we have
 \beq\label{vv00242}
  \begin{array}{c}
  \displaystyle{
 r_{12}(z)=\sum\limits_{b}\frac{\p l(z)}{\p
 S_b}\otimes T_{-b}\,.
 }
 \end{array}
 \eq
This relation was used in \cite{AASZ} for computation of the
rational classical $r$-matrix.

\subsection{$\eta$-independent quadratic Poisson
brackets}\label{quad2}



Let us consider another limit of brackets  (\ref{vv019}) and
(\ref{vv021}). Set
 \beq\label{vv030}
 \begin{array}{c}
  \displaystyle{
{\mathcal S}_0=\eta\,{\mathrm s}_0\,.
 }
 \end{array}
 \eq
With this rescaling ${\mathcal S}_0\rightarrow 0$ when
$\eta\rightarrow 0$. Then the residue  $S$ becomes traceless, i.e.
 \beq\label{vv026}
 \begin{array}{c}
  \displaystyle{
S\ \stackrel{\eta\rightarrow 0}{\longrightarrow}\ \bar
S=S-\frac{1}{N}\,\tr(S)\,1_{N\times N}\,,
 }
 \end{array}
 \eq
  \beq\label{vv025}
 \begin{array}{c}
  \displaystyle{
l(z)\ \stackrel{\eta\rightarrow 0}{\longrightarrow}\ \bar
l(z)=l(z)-\frac{1}{Nz}\,\tr (S)\,1_{N\times N}=\frac{1}{z}\bar S+
l^{(0)}(S)+O(z)\,.
 }
 \end{array}
 \eq
Applying this limit to
(\ref{vv019}) we get
 \beq\label{vv031}
 \begin{array}{c}
  \displaystyle{
 \{{\mathrm L}_1(z)\,, {\mathrm L}_2(w)\}=[\, {\mathrm L}_1(z)\, {\mathrm L}_2(w),r_{12}(z-w)]\,,
 }
 \end{array}
 \eq
 where the Lax matrix
 \beq\label{vv0311}
 \begin{array}{c}
  \displaystyle{
{\mathrm L}(z):=\frac{\mathrm s_0}{N}\,1+\bar l(z)\,.
 }
 \end{array}
 \eq
The Poisson brackets (\ref{vv021}) acquire the following form in the
limit:
 \beq\label{vv032}
 \begin{array}{c}
  \displaystyle{
{\mathcal A}_{\hbar=0,\eta=0}^{\hbox{\tiny{Skl}}}:\ \ \ \{{\bar
S}_1,{\bar S}_2\}=\frac{\mathrm s_0}{N}\,[{\bar S}_2,P_{12}] +[{\bar
S}_1{\bar S}_2,r_{12}^{(0)}]+[\tr_3
 (r_{13}^{(0)}{\bar S}_3)\,{\bar S}_2,P_{12}]\,.
 }
 \end{array}
 \eq
The missing brackets $\{\bar S,\mathrm s_0\}$ can be found by taking
the limit in (\ref{vv0142}). Plugging (\ref{vv030}) into
(\ref{vv0142}), (\ref{vv0143}) we get:
  \beq\label{vv033}
  \begin{array}{c}
  \displaystyle{
\{{\mathrm s}_0,\bar S\}=\lim\limits_{\eta\rightarrow 0}\frac{[S,{
J^\eta}( S)]}{\eta}=[\bar S,{\mathrm J}(\bar S)]\,,
 }
 \end{array}
 \eq
where the inverse inertia tensor $\mathrm J$ is defined as
   \beq\label{vv034}
  \begin{array}{c}
  \displaystyle{
 \mathrm J:\ \ \bar S_1\ \rightarrow\ \mathrm J(\bar S)_1=\tr_2\left(r'_{12}(0)\bar S_2\right)
   }
   \end{array}
  \eq
with $r'_{12}(0)$ from (\ref{vv018}). Equations (\ref{vv033}),
(\ref{vv034}) also play the role of equations of motion generated by
the Hamiltonian ${\mathrm s}_0$:
  \beq\label{vv0332}
  \begin{array}{c}
  \displaystyle{
\p_{\mathrm t_0}\,\bar S= \{{\mathrm s}_0,\bar S\}=[\bar S,{\mathrm
J}(\bar S)]\,,
 }
 \end{array}
 \eq
The corresponding Lax equation can be obtained in two ways. The
first one \cite{Babel} -- is by taking $\tr_2$ in (\ref{vv031}).
This yields
  \beq\label{vv035}
  \begin{array}{c}
  \displaystyle{
\p_{\mathrm t_0}\, \mathrm L(z)=\{{\mathrm s}_0,\mathrm
L(z)\}=[\mathrm L(z)\,,\mathrm M(z)]\,,
 }
 \end{array}
 \eq
where
  \beq\label{vv036}
  \begin{array}{c}
  \displaystyle{
\mathrm M_1(z)=-\tr_2\left(r_{12}(z-w)\mathrm L_2(w)\right)\,.
 }
 \end{array}
 \eq
The latter matrix should be $w$-independent up to some element from
the kernel of $\hbox{ad}_{\mathrm L(z)}$. Alternatively, one can
consider the limit of (\ref{vv015}): $\{{\mathcal S}_0\,,L^\eta
(z)\}=[ L^\eta(z)\,, M(z)]$. Notice again that $M(z)$ given by
(\ref{vv016}) is $\eta$-independent. Moreover,  it coincides with
${\mathrm L(z)}$ up to sign and some scalar - element from
$\hbox{Ker}(\hbox{ad}_{\mathrm L(z)})$, i.e.
  \beq\label{vv037}
  \begin{array}{c}
  \displaystyle{
 \{{\mathcal S}_0\,,L^\eta
(z)\}=[ L^\eta(z)\,, M(z)]=-[ L^\eta(z)\,, \mathrm L(z)]\,.
 }
 \end{array}
 \eq
Then, substituting rescaling (\ref{vv030}) and using expansion
(\ref{vv017}) we get
  \beq\label{vv038}
  \begin{array}{c}
  \displaystyle{
\p_{\mathrm t_0}\, \mathrm L(z)=\{{\mathrm s}_0\,,\mathrm
L(z)\}=\lim\limits_{\eta\rightarrow 0}\frac{[\mathrm L(z)\,,
 L^\eta(z)]}{\eta}=[\mathrm L(z)\,,{\mathcal M}(z)]\,,
 }
 \end{array}
 \eq
where
  \beq\label{vv039}
  \begin{array}{c}
  \displaystyle{
{\mathcal M}(z)=\tr_2\left(r'_{12}(z)\bar S_2\right)\,.
 }
 \end{array}
 \eq
 with $r'_{12}(z)$ defined in (\ref{vv017}).
Thus, the roles of $L$ and $M$-operators are interchanged while
taking the limit. In addition, $r'_{12}(z)$ has no singularities at
$z=0$. Then
  \beq\label{vv0392}
  \begin{array}{c}
  \displaystyle{
{\mathcal M}(0)=\tr_2\left(r'_{12}(0)\bar
S_2\right)\stackrel{(\ref{vv034})}{=}\mathrm J(\bar S)\,.
 }
 \end{array}
 \eq
Finally, we see that the expansion (\ref{vv017}) of $L^\eta(z)$ in
$\eta$ provides $M$-operators for both -- $\eta$-dependent and
$\eta$-independent descriptions:
  \beq\label{vv0393}
  \begin{array}{c}
  \displaystyle{
L^\eta(z)=\eta^{-1}{\mathcal S}_0/N-M(z) +\eta{\mathcal
M}(z)+O(\eta^2)\,.
 }
 \end{array}
 \eq
Notice also that the $M$-operator (\ref{vv039}) is also valid for
the linear $r$-matrix structure (\ref{vv023}) since the Lax pairs
for the linear and quadratic ($\eta$-independent) $r$-matrix
structures are the same (up to scalar terms). The formulae obtained
in this section can be considered as an extension of \cite{Babel}
for the class of integrable systems under consideration.

\noindent {\bf Relation between ${\mathcal A}_{\eta\neq
0}^{\hbox{\tiny{Skl}}}$ and ${\mathcal
A}_{\eta=0}^{\hbox{\tiny{Skl}}}$.} We have two different description
of the classical Sklyanin algebras (and related integrable models).
In the first one the quadratic Poisson structure (\ref{vv009})
${\mathcal A}_{\hbar=0,\eta}^{\hbox{\tiny{Skl}}}$ is
$\eta$-dependent. The second  ${\mathcal
A}_{\hbar=0,\eta=0}^{\hbox{\tiny{Skl}}}$ (\ref{vv032}),
(\ref{vv033}) -- is $\eta$-independent. The same happens at quantum
level. One can quantize the Lax matrix (\ref{vv0311}) as
 \beq\label{vv03117}
 \begin{array}{c}
  \displaystyle{
\hat{\mathrm L}(z):=\frac{1}{N}\,\hat\mathrm s_0+\tr_2\left(
r_{12}(z){\hat{\bar{S}}}_2\right)\,.
 }
 \end{array}
 \eq
Then the exchange relations (\ref{vv001}) gives  $\eta$-independent
quadratic algebra ${\mathcal A}_{\hbar,\eta=0}^{\hbox{\tiny{Skl}}}$.

Algebras ${\mathcal A}_{\eta\neq 0}^{\hbox{\tiny{Skl}}}$ and
${\mathcal A}_{\eta=0}^{\hbox{\tiny{Skl}}}$ are related. The
relation is easy to demonstrate explicitly in the elliptic case (see
Section \ref{ell}). The idea is the following. There exists a linear
functional $\vf^\eta$ on ${\rm gl}_N$ (depending on the boundary
conditions) such that (\ref{vv0025}) and (\ref{vv03117}) are related
as follows:
 \beq\label{vv0314}
 \begin{array}{c}
  \displaystyle{
g^\eta(z)\,e^{\eta_0\p_z}\hat L(z,\vf^\eta(\hat S))=\hat{\mathrm
L}(z, \hat S)\,,
 }
 \end{array}
 \eq
where $g^\eta(z)$ is some function with a simple zero at $z=-\eta_0$
and simple pole at $z=0$. Finally, the relation can be written as
 \beq\label{vv03149}
 \begin{array}{c}
  \displaystyle{
 L^\eta\Big(z+\eta_0,{\mathrm
L}(-\eta_0,S)\Big)=\frac{\tr L^\eta\left(z+\eta_0,S\right)}{\tr
S}\,{\mathrm L}(z,S)\,.
 }
 \end{array}
 \eq
%
%
It holds true in the rational case as well. We may use this relation
to get explicit change of variables from the $\eta$-independent
description (\ref{vv0311}) to the $\eta$-dependent (\ref{vv0084}):
 \beq\label{vv71165}
 \begin{array}{c}
  \displaystyle{
 S\ \to\ c\,\mathrm L(c\eta,S)\,,\ \ c=-\frac{\eta_0}{\eta}\,.
 }
  \end{array}
 \eq
The coefficient $-{\eta_0}/{\eta}$ is chosen in order to have
$\res\limits_{\eta=0}\left(-\frac{\eta_0}{\eta}\,\mathrm
L(-\eta_0,S)\right)=\res\limits_{z=0}\mathrm (z,S)$.     It is
interesting to note that plugging this change of variables to the
equations of motion (\ref{vv0142}) gives
 \beq\label{vv71168}
 \begin{array}{c}
  \displaystyle{
 \p_t\mathrm L(c\eta,S)=[\mathrm L(c\eta,S),cJ^\eta(\mathrm
 L(c\eta,S))]\,,
 }
  \end{array}
 \eq
 i.e. the Lax equations
(\ref{vv035}), where $c\eta$ plays the role of the spectral
parameter. Hence, we get an alternative definition for the
$M$-operator
 \beq\label{vv71169}
 \begin{array}{c}
  \displaystyle{
 \mathrm M(z,S)=c\, J^{z/c}(\mathrm  L(z,S))\,.
 }
  \end{array}
 \eq

\section{Relativistic rational top}\label{rat}
\setcounter{equation}{0}

In this section we obtain explicit answer for the Lax pair of the
relativistic top. As it was already mentioned this model is a
top-like form of the spin Ruijsenaars-Schneider (RS) model. To get
the answer we represent the Lax matrix of RS model in the factorized
form (\ref{v063}) which is convenient for the gauge transformation.
The dynamical variables of the top are the components of the residue
(\ref{vv507}), (\ref{vv515}) of the gauge transformed RS Lax matrix
(\ref{vv506}). We express the gauge transformed $L$-operator in
terms of its residue. This gives the correct answer for generic top
since it is independent of the Casimir functions values.

\subsection{Factorized $L$-operators for classical Ruijsenaars-Schneider model
}\label{Ruijs} 

In this paragraph we propose factorized forms of $L$-operators for
the rational RS model  \cite{Ruijs1}.


Following \cite{AASZ} for the set of variables $\{q_j\}$, $j=1...N$
such that $\sum\limits_{j=1}^N q_j=0$
 let us introduce the matrix
   \beq\label{vv417}
  \begin{array}{c}
  \displaystyle{
\Xi_{ij}(\bfq,z):=(z+q_j)^{\varrho(i)}\,,\ \ i,j=1...N\,,
   }
   \end{array}
  \eq
 where
  \beq\label{vv922}
\varrho(i)=\left\{\begin{array}{ll}
i-1 & {\rm{for}}\ \ 1\leq i\leq N-1,\\
 & \\
i & {\rm{for}}\ \ i= N.
\end{array}\right.\
 \hskip10mm
 \varrho^{-1}(i)=\left\{\begin{array}{ll}
i+1 & {\rm{for}}\ \ 0\leq i\leq N-2,\\
 & \\
i & {\rm{for}}\ \ i= N.
\end{array}\right.
 \eq
It has the property
 \beq\label{vv911}
 \begin{array}{c}
 \displaystyle{
\det\Xi(\bfq,z)=Nz \prod\limits_{1\geq i>j\geq N}(q_i-q_j)\,,
 }
 \end{array}
 \eq
i.e. the matrix is degenerated at $z=0$. It can be also treated as
the rational analogue of the modification of bundles over elliptic
curves used in \cite{LOZ} for the description of the elliptic top.

\vspace{4mm}

\noindent{\bf Rational ${\rm sl}_N$ RS model with spectral
parameter} is defined by the following Lax matrix:
 \beq\label{v061}
 {L}^{\hbox{\tiny{RS}}}_{ij}(z,\eta)=\eta\,
 \left(\frac{1}{q_i-q_j+\eta}-\frac{1}{Nz} \right)\,e^{ p_j/c}\,\prod\limits_{k\neq
j}^N\frac{q_j-q_k-\eta}{q_j-q_k}\,.
  \eq
  where
   \beq\label{v06177}
\sum\limits_{k=1}^N q_k=\sum\limits_{k=1}^N p_k=0\,.
  \eq
The classical $r$-matrix structure was found in \cite{Avan}.
 \begin{predl}
The Lax matrix (\ref{v061}) can be written in the following
form\footnote{The prove is direct. See formulae in Section \ref{lax}
and Appendix in \cite{AASZ}.}:
 \beq\label{v063}
 \begin{array}{c}
 {L}^{\hbox{\tiny{RS}}}(z,\eta)=\,D_0({\bf q})\, \Xi^{-1}({\bf q}, z)\,
 \Xi({\bf q}, z-\eta)\,D_0^{-1}({\bf q})\,e^{P/c}\,,
 \end{array}
  \eq
where $D_0({\bf q})$ is diagonal matrix
$\left(D_0\right)_{ij}=\delta_{ij}\prod\limits_{k\neq
 i}^N(q_i-q_k)$.
 \end{predl}

\vspace{2mm}

\noindent 
Let us also write the similar answer for\footnote{It is not used in
the subsequent sections.}

\noindent{\bf Rational ${\rm sl}_N$ RS model without spectral
parameter.} The Lax matrix
 \beq\label{v051}
 {L}^{\hbox{\tiny{RS}}}_{ij}={{\eta\, e^{p_j/c}}
 \over{q_i-q_j+\eta}}\prod\limits_{k\neq
j}^N\frac{q_j-q_k-\eta}{q_j-q_k}\,.
  \eq
  is represented in the form:
 \beq\label{v053}
 \begin{array}{c}
 {L}^{\hbox{\tiny{RS}}}=\,D_0({\bf q})\, V^{-1}({\bf q}, z)\,
 V({\bf q}, z-\eta)\,D_0^{-1}({\bf q})\,e^{P/c}=\\ \ \\
=\,D_0({\bf q})\, V^{-1}({\bf q}, z)\,C_{-\eta}\,V({\bf q},
z)\,D_0^{-1}({\bf q})\,e^{ P/c}\,,
 \end{array}
  \eq
where
   \beq\label{vv4175}
  \begin{array}{c}
  \displaystyle{
V_{ij}(\bfq,z):=(z+q_j)^{i-1}\,,\ \ i,j=1...N\,,
   }
   \end{array}
  \eq
  and
 \beq\label{v054}
 \begin{array}{c}
{\left(C_{\lambda}\right)}_{ij}=
\left\{\begin{array}{l}\displaystyle{
\frac{(i-1)!\, \lambda^{i-j} }{(j-1)!(i-j)!}\,,}\ \ j\leq i\,,\\ \ \\
0\,,\ \ j>i\,.
\end{array}\right.
\end{array}
  \eq
 It easy to verify that
  \beq\label{v0551}
C_\lambda=\exp(\lambda C_0)\,,\ \ \ (C_{0})_{ij}=
\left\{\begin{array}{l} j\,,\ \ \ i=j+1\,,\ i=2\,,...\,,N,\\  0\,,\
\ \  {otherwise}
\end{array}\right.
  \eq
  and $\p_zV=C_0V$.
The limit to Calogero-Moser model is obtained as follows:
 \beq\label{v055}
L^{\hbox{\tiny{CM}}}_{ij}= \lim\limits_{c\rightarrow \infty}
c\,{L^{\hbox{\tiny{RS}}}_{ij}\left.\right|_{\,\eta=\nu/c}-\delta_{ij}}=P-\nu
D_0V^{-1} C_0 V
 D_0^{-1}\,.
  \eq
Notice that we can also define the Lax matrix as
 \beq\label{v056}
 {L'}^{\,\hbox{\tiny{RS}}}_{ij}={{\eta\, e^{p_j/c}}
 \over{q_i-q_j+\eta}}\prod\limits_{k\neq
j}^N\frac{q_j-q_k+\eta}{q_j-q_k}\,.
  \eq
It differs from (\ref{v051}) by the canonical  map
 \beq\label{v052}
e^{ p_j/c}\ \longrightarrow\ e^{ p_j/c} \prod\limits_{k\neq
j}\left(\frac{q_j-q_k+\xi}{q_j-q_k-\xi}\right)^a
  \eq
with $a=1$ and $\xi=\eta$. Then
 \beq\label{v057}
 \begin{array}{c}
 {L'}^{\,\hbox{\tiny{RS}}}=
 \,D^{-1}_{\eta}({\bf q})\,\left(V^T\right)({\bf q},z+\eta)\,\left(V^T\right)^{-1}({\bf q}, z)\,
 D_{\eta}({\bf q})\,e^{P/c}
 =\\ \ \\
 =\,D^{-1}_{\eta}({\bf q})\,\left(V^T\right)({\bf q}, z)\,
 C^T_{\eta}\,\left(V^T\right)^{-1}({\bf q}, z)\,D_{\eta}({\bf
 q})\,e^{P/c}\,,\ \ \ \left(D_\lambda\right)_{ij}=\delta_{ij}\prod\limits_{k\neq
 i}^N(q_i-q_k+\lambda).
 \end{array}
  \eq
Let us mention that the transformation (\ref{vv4175}) was used in
the classical \cite{Feher1} and the quantum
\cite{Hodges1} IRF-Vertex transformations. In this way the
(Jordanian) $R$-matrices of the Cremmer-Gervais type were obtained.

\subsection{Lax pair}\label{lax}

Apply the gauge transformation $g(z)=\Xi({\bf q}, z)\,D_0^{-1}({\bf
q})$ to the RS Lax matrix (\ref{v061}), (\ref{v063}) with
$\eta:=-\eta$. Then it follows from (\ref{v063}) that
 \beq\label{vv506}
  \begin{array}{c}
  \displaystyle{
\tilde L^{\eta,c}(z,\bfq,{\bf p}):=\Xi({\bf q}, z)\,D_0^{-1}({\bf
q})\,{L}^{\hbox{\tiny{RS}}}_{ij}(z,-\eta)\,D_0({\bf
q})\,\Xi^{-1}({\bf q}, z)=}\\ \ \\
 \displaystyle{
 =\Xi({\bf q}, z+\eta)\,e^{P/c}\,\Xi^{-1}({\bf q}, z)\,.
   }
   \end{array}
  \eq
Let $\sum\limits_{j=1}^N p_j=0$. Then from (\ref{vv911}) it follows
that
 \beq\label{vv5062}
  \begin{array}{c}
  \displaystyle{
\det\tilde L^{\eta,c}(z,\bfq,{\bf p})=\frac{z+\eta}{z}\,.
   }
   \end{array}
  \eq
Our purpose now is to express this matrix in terms of its residue at
$z=0$. Set
 \beq\label{vv507}
  \begin{array}{c}
  \displaystyle{
S=N\res\limits_{z=0}\tilde L^{\eta,c}(z,\bfq,{\bf p})=
\frac{1}{\prod\limits_{1\geq i>j\geq N}(q_i-q_j)}\, \Xi({\bf q},
\eta)\,e^{P/c}\,{\bf \hbox{adj}}\,\left(\Xi({\bf q}, 0)\right)\,,
   }
   \end{array}
  \eq
where the adjugate is  transpose of the cofactor matrix. To find
matrix components we need the inverse of $\Xi$:
 \beq\label{vv5075}
 \begin{array}{c}
 \displaystyle{
\Xi^{-1}_{k
j}({\bfx})=(-1)^{\varrho(j)}\,\frac{\sigma_{\varrho(j)}({\bfx})}{(\sum\limits_{s=1}^{N}
\,x_{s})\prod\limits_{s\neq k}^N
(x_{k}-x_{s})}-(-1)^{\varrho(j)}\,\frac{\stackrel{k}{\sigma}_{\varrho(j)}({\bfx})}{\prod\limits_{s\neq
k}^{N} (x_{k}-x_{s})} }\,,
 \end{array}
 \eq
where $x_j=q_j+z$. Expansion in powers of $z$ gives
 \beq\label{vv508}
 \begin{array}{c}
 \displaystyle{
 \Xi^{-1}_{mj}(z,{\bf
 q})=\frac{1}{Nz}\frac{(-1)^{\varrho(j)}}{\prod\limits_{r\neq
 m}^N(q_m-q_r) }\Big(
 \sigma_{\varrho(j)}({\bfq})+\sum\limits_{s=1}^{N-j} z^s\left[
 \sigma_{s+j-1}(\bfq) \left(\!\begin{array}{c} s+j-1\\
 j-1\end{array}\!\right) \right.}
\\
\
\\
\displaystyle{ \left.
-N \stackrel{m}{\sigma}_{s+j-2}\!({\bfq}) \left(\!\begin{array}{c} s+j-2\\
 j-1\end{array}\!\right)
 \right] -(N-j)\,z^{N-j+1} \stackrel{m}{\sigma}_{N-1}\!({\bfq})  \left(\!\begin{array}{c} N\\
 j-1\end{array}\!\right) \Big)\,.
 }
 \end{array}
 \eq
In (\ref{vv5075}) and (\ref{vv508}) the elementary symmetric
functions are used:
 \beq\label{vv510}
 \begin{array}{c}
 \displaystyle{
\prod\limits_{k=1}^{N} \,(\zeta-x_{k})= \sum\limits_{k=0}^{N}
(-1)^{k} \zeta^{k} \sigma_{k}({\bf x})
 }
 \end{array}
 \eq
or
 \beq\label{vv511}
 \begin{array}{c}
 \displaystyle{
\sigma_{N-d}(\bfx)=(-1)^N\sum\limits_{1 \leq i_{1} <
i_{2}...<i_{d}\leq N} x_{i_{1}} x_{i_{2}}...x_{i_{d}}\,,\ \ \
d=0,...,N
 }
 \end{array}
 \eq
 and their derivatives
 \beq\label{vv512}
 \begin{array}{c}
 \displaystyle{
-\prod\limits_{m\neq k}^{N} \,(\zeta-x_{m})=\sum\limits_{s=0}^{N-1}
(-1)^{s} \zeta^{s} \stackrel{k}{\sigma}_s(\bfx)\,.
 }
 \end{array}
 \eq
These functions satisfy the following set of identities:
 \beq\label{vv513}
 \begin{array}{c}
 \displaystyle{
\stackrel{m}{\sigma}_{j}({\bfx})=\sum\limits_{c=0}^{N-j-1}
(-x_m)^{c} \sigma_{j+1+c}(\bfx)\,,
 }
 \end{array}
 \eq
 \beq\label{vv514}
 \begin{array}{c}
 \displaystyle{
\stackrel{m}{\sigma}_{j}({\bfx})=-\sum\limits_{c=0}^{j}
(-x_m)^{-1-c} \sigma_{j-c}(\bfx)\,.
 }
 \end{array}
 \eq
From (\ref{vv5075}) we can easily find $S$ from (\ref{vv507}):
 \beq\label{vv515}
  \begin{array}{|c|}
  \hline\\
 \displaystyle{
S_{ij}={N}\res\limits_{z=0}L^{\hbox{\tiny{top}}}_{ij}(z)=\sum_{m=1}^{N}\,\frac{({
q}_{m}+\eta)^{\,\varrho(i)} e^{p_{m}/c} }{ \prod\limits_{k\neq
m}^{\,} ({ q}_{m}-{
q}_{k})}\,\,\,(-1)^{\varrho(j)}\,\sigma_{\varrho(j)}(\bfq)
 }
   \\ \ \\
 \hline
 \end{array}
 \eq
 To take into account the light speed $c$ in the quadratic brackets
 (\ref{vv009}) one should put the common factor $1/c$ in front of
 r.h.s. of  (\ref{vv009}). It is equivalent to redefinition of the
 classical $r$-matrix $r_{12}(z)\to r_{12}(z)/c$.

Using (\ref{vv508})-(\ref{vv514}) we can rewrite the Lax matrix
(\ref{vv506}) in terms of the variables $S$ (\ref{vv508}). The
computation gives:
{\small{
 \beq\label{vv352}
 \begin{array}{c}
  \displaystyle{
{ L}^{\eta}(z)=N\,\tilde L^{\eta,c}(z,\bfq,{\bf
p})=\frac{1}{z}\sum\limits_{i,j=1}^N
 {\mathrm E}_{ij}\
 \left\{\,\,\sum\limits_{\ga=0}^{\varrho(i)} z^\gamma \left(\!\begin{array}{c} \varrho(i) \\ \gamma
 \end{array}\!\right)
 S_{\varrho^{-1}(\varrho(i)-\ga),j}\right.
}
 \\
 \displaystyle{
  -\sum\limits_{\ga=0}^{\varrho(i)} z^{\ga+N-j+1}\,
 (-1)^{\varrho(j)+N}(N\!-\!j)\left(\!\begin{array}{c} \varrho(i) \\ \gamma
 \end{array}\!\right) \left(\!\begin{array}{c} N \\ j\!-\!1
 \end{array}\!\right) S_{\varrho^{-1}(\varrho(i)-\ga),\,N}
 }
\\
  \displaystyle{
+\!\sum\limits_{\ga=0}^{\varrho(i)} \,\sum\limits_{s=1}^{N-j}
z^{s+\ga}\, (-1)^{\varrho(j)+s+j-1} \!\left(\!\begin{array}{c} \varrho(i) \\
\gamma
 \end{array}\!\right) \left(\!\begin{array}{c} s\!+\!j\!-\!1 \\ j\!-\!1
 \end{array}\!\right)
 S_{\varrho^{-1}(\varrho(i)-\ga),\,\varrho^{-1}(s+j-1)}\,-
 }
 \end{array}
 \eq
 \beq\label{vv353}
 \begin{array}{c}
 \displaystyle{
  -N\sum\limits_{s=1}^{N-j}\, \sum\limits_{b=0}^{\varrho(i)}
  (-1)^{\varrho(j)\!+s+j-1}\, z^s\, (z+\eta)^b \left(\!\begin{array}{c} s\!+\!j\!-\!2 \\ j\!-\!1
 \end{array}\!\right) \left(\!\begin{array}{c} \varrho(i) \\ b
 \end{array}\!\right)\times
  }
   \\
 \displaystyle{\left[
 \delta_{\varrho(i)-j-s-b+1\leq\, 0} \sum\limits_{c=0}^{N\!-s-j+1}\ \sum\limits_{p=0}^{\varrho(i)\!-b+c}
   (-\eta)^p\left(\!\begin{array}{c} \varrho(i)\!-\!b\!+\!c \\ p
 \end{array}\!\right)  S_{\varrho^{-1}(\varrho(i)-b-p+c),\varrho^{-1}(s+j+c-1)}
  \right.}
     \\
 \displaystyle{\left.
\left. -\delta_{\varrho(i)-j-s-b+1>\, 0} \sum\limits_{c=0}^{s+j-2}\
\sum\limits_{p=0}^{\varrho(i)\!-b-c-1}
   (-\eta)^p\left(\!\begin{array}{c} \varrho(i)\!-\!b\!-\!c\!-\!1 \\ p
 \end{array}\!\right)  S_{\varrho^{-1}(\varrho(i)-b-p-c-1),\varrho^{-1}(s+j-c-2)}
  \right]\,\right\}}
 \end{array}
 \eq
 \beq\label{vv354}
 \begin{array}{c}
  \displaystyle{
+\frac{1}{z}\left[zN\,{\mathrm E}_{NN}-\sum\limits_{j=1}^N
z^{N-j+2}(-1)^{\varrho(j)+N} (N-j)N \left(\!\begin{array}{c} N \\
j-1 \end{array}\!\right) {\mathrm E}_{Nj}\right.
  }
  \\
   \displaystyle{
  -N\sum\limits_{i,j=1}^N\,\sum\limits_{s=1}^{N-j}\, \sum\limits_{b=0}^{\varrho(i)}
  (-1)^{\varrho(j)\!+s+j-1}\, z^s\, (z+\eta)^b \left(\!\begin{array}{c} s\!+\!j\!-\!2 \\ j\!-\!1
 \end{array}\!\right) \left(\!\begin{array}{c} \varrho(i) \\ b
 \end{array}\!\right)\delta_{p\leq 1}\,\delta_{\varrho(i)-b-p-j-s+2\,,\,0}\,\times
  }
     \\
 \displaystyle{
 \left.\sum\limits_{p=0}^{\varrho(i)-b+N-s-j+1}
   (-\eta)^p\left(\!\begin{array}{c} \varrho(i)-b+N\!-s-j+1 \\ p
 \end{array}\!\right)\, {\mathrm E}_{ij}\,  \right] \times
 }
 \end{array}
 \eq
 \beq\label{vv356}
 \begin{array}{c}
  \displaystyle{
\times\left(-\frac{1}{N(-\eta)}\sum\limits_{j=1}^N
S_{jj}-\frac{1}{N^2}\sum\limits_{j=1}^N\Big[\delta_{\varrho(j)\geq
1} \,\varrho(j)\,
S_{\varrho^{-1}(\varrho(j)-1),\,\varrho^{-1}(j)}+(-1)^{\varrho(j)+j}\,
j\,\, S_{j,\varrho^{-1}(j)}\right.
  }
  \\
  \displaystyle{\left.
  -N\sum\limits_{b=0}^{\varrho(j)} (-1)^{\varrho(j)\!+\!j+b}\, \left(\!\begin{array}{c} \varrho(j) \\ b
 \end{array}\!\right)
 \sum\limits_{c=0}^{N-j}\ \sum\limits_{p=0}^{\varrho(j)\!-b+c}
   (-\eta)^{p+b}\left(\!\begin{array}{c} \varrho(j)\!-\!b\!+\!c \\ p
 \end{array}\!\right)
 S_{\varrho^{-1}(\varrho(j)-b-p+c),\varrho^{-1}(j+c)}\Big]\right)
  }
 \end{array}
 \eq
}}
Here and below we imply that the values of indices corresponding to
the undefined argument value ($N-1$) of function $\varrho^{-1}$
(\ref{vv922}) are skipped in the summations.

It is  important to mention that the answer does not depend on the
values of the Casimir functions (\ref{vv011}). As we know it is
defined by only quantum $R$-matrix which is the subject of the next
Section. Therefore, we can consider the obtained expression
(\ref{vv352})-(\ref{vv356}) as independent definition of the Lax
matrix for the generic {\em rational relativistic top}. 
The case (\ref{vv506}) which is gauge equivalent
 to the rational RS model appears for the particular values of the
 Casimir functions $C_k$ (\ref{vv011}). From (\ref{vv5062}) we
 conclude that the RS case corresponds to:
 \beq\label{vv380}
 \begin{array}{c}
  \displaystyle{
 \hbox{RS}:\ \ \ C_{-1}=\eta\,,\ \ C_0=1\,,\ \ C_k=0\,, k\neq
 -1\,,0\,.
 }
 \end{array}
 \eq
In the non-relativistic limit this case corresponds to the rational
top on the coadjoint orbit of minimal dimension ($2N-2$). In a
general case the obtained model yields alternative description of
the spin RS model \cite{KrichZabr}.

The quantum Lax matrix is obtained from (\ref{vv352})-(\ref{vv356})
by substitution $S\rightarrow\hat S$. In (\ref{vv515}) it
corresponds to $p_j:=\hbar\p_{q_j}$ with the choice of normal
ordering.

\vskip3mm

\noindent {\bf M-operator from non-relativistic limit.}
As it was shown in (\ref{vv020}) the $M$-operator (\ref{vv016})
coincides (up to minus) with the non-relativistic limit of the Lax
matrix. Hence, we can use the answer obtained in \cite{AASZ}:
\footnote{Notice that in \cite{AASZ} the answer is ${\rm
sl}_N$-valued. It differs from (\ref{vv020}) by scalar matrix and
factor $-N$.}:
{\small{
 \beq\label{vv381}
 {M}_{ij}(z)=-\frac{1}{z}\times
 \eq
 $$
\begin{array}{c}
 \displaystyle{
  \Big[\sum\limits_{\ga=0}^{\varrho(i)} z^\gamma \left(\!\begin{array}{c} \varrho(i) \\ \gamma
 \end{array}\!\right)
 S_{\varrho^{-1}(\varrho(i)-\ga),j}-\sum\limits_{\ga=0}^{\varrho(i)} z^{\ga+N-j+1}\,
 (-1)^{\varrho(j)+N}(N\!-\!j)\left(\!\begin{array}{c} \varrho(i) \\ \gamma
 \end{array}\!\right) \left(\!\begin{array}{c} N \\ j\!-\!1
 \end{array}\!\right) S_{\varrho^{-1}(\varrho(i)-\ga),\,N}
 }
 \\
  \displaystyle{
-N\sum\limits_{\ga=0}^{\varrho(i)} \sum\limits_{s=1}^{N-j}
\delta_{\varrho(i)-j+1\leq s+\ga} z^{s+\ga}\,
(-1)^{\varrho(j)+s+j-1}\left(\!\begin{array}{c} \varrho(i) \\
\gamma
 \end{array}\!\right) \left(\!\begin{array}{c} s\!+\!j\!-\!2 \\ j\!-\!1
 \end{array}\!\right)\sum\limits_{c=0}^{N-s-j+1} S_{\varrho^{-1}(\varrho(i)\!-\!\ga\!+\!c),\,\varrho^{-1}(s\!+\!j\!+\!c\!-\!1)}
 }
  \\
  \displaystyle{
-N\sum\limits_{\ga=0}^{\varrho(i)} \sum\limits_{s=1}^{N-j}
\delta_{\varrho(i)-j+1> s+\ga} z^{s+\ga}\,
(-1)^{\varrho(j)+s+j-2}\left(\!\begin{array}{c} \varrho(i) \\
\gamma
 \end{array}\!\right) \left(\!\begin{array}{c} s\!+\!j\!-\!2 \\ j\!-\!1
 \end{array}\!\right)\sum\limits_{c=0}^{s+j-2} S_{\varrho^{-1}(\varrho(i)\!-\!\ga\!-\!c\!-1),\,\varrho^{-1}(s\!+\!j\!-\!c\!-2)}
 }
 \\
  \displaystyle{
+\!\sum\limits_{\ga=0}^{\varrho(i)} \sum\limits_{s=1}^{N-j}
z^{s+\ga}\, (-1)^{\varrho(j)+s+j-1} \!\left(\!\begin{array}{c} \varrho(i) \\
\gamma
 \end{array}\!\right) \left(\!\begin{array}{c} s\!+\!j\!-\!1 \\ j\!-\!1
 \end{array}\!\right) S_{\varrho^{-1}(\varrho(i)-\ga),\,\varrho^{-1}(s+j-1)}\!-\!\frac{\delta_{i,j}}{N}
\sum\limits_{k=1}^N\sum\limits_{c=0}^{N\!-\!k\!-2}S_{k+c,\,k+c+1}\Big]
 }
 \end{array}
 $$
}}

\vskip3mm

\noindent {\bf Example:} for $N=2$ (\ref{vv352})-(\ref{vv356})
yields the Lax matrix
 \beq\label{vv360}
 \begin{array}{c}
  \displaystyle{
{ L}^{\eta}(z)=\frac{1}{z}S_{2\times
2}+\frac{\tr(S)}{\eta}1_{2\times 2} -(z+\eta)
 \left(
 \begin{array}{cc}
{ S_{12}}&{0}
\\ \ \\
{(S_{11}-S_{22}) +(\eta^2+z^2+\eta z)S_{12}}&{ -S_{12}}
 \end{array}
 \right)
  }
 \end{array}
 \eq
 with $S_{2\times
2}=\mat{S_{11}}{S_{12}}{S_{21}}{S_{22}}$ and $\tr S=S_{11}+S_{22}$.
The determinant defines the Casimir functions
 \beq\label{vv3607}
 \begin{array}{c}
  \displaystyle{
\det{
L}^{\eta}(z)=\frac{1}{z^2}C_0+(\frac{1}{z\eta}+\frac{1}{\eta^2})C_1\,,
  }
 \end{array}
 \eq
 $$
C_0=\det S=S_{11}S_{22}-S_{12}S_{21}\,,\ \ \
C_1=(S_{11}+S_{22}+\eta^2 S_{12})^2-4\eta^2 S_{12}S_{22}
 $$
 of the Poisson structure (\ref{vv009}) for ${\rm gl}_2$.
The Hamiltonian $S_{11}+S_{22}$ generates equations
 of motion (\ref{vv0142}) with the $M$-operator (\ref{vv016}):
 \beq\label{vv3663}
 \begin{array}{c}
  \displaystyle{
M(z)=  -\frac{1}{z}\left( \begin{array}{cc} S_{11}-z^2S_{12} &
S_{12}
\\ \ \\ 
S_{21}-z^2 (S_{11}-S_{22})-z^4S_{12} & S_{22}+z^2S_{12}
\end{array} \right)
  }
 \end{array}
 \eq
In $N=2$ case $r_{12}^{(0)}$  from (\ref{vv0031}) vanishes. Hence,
$J^\eta(S)$ (\ref{vv0143}) is defined by only $L^{\eta,(0)}$ from
(\ref{vv0021}):
 \beq\label{vv3673}
 \begin{array}{c}
  \displaystyle{
J^\eta(S)=L^{\eta,(0)}=-\mat{\eta S_{12}}{0}{\eta^3 S_{12}+\eta
(S_{11}-S_{22})}{-\eta S_{12}} +
\frac{S_{11}+S_{22}}{\eta}\,1_{2\times 2}\,.
  }
 \end{array}
 \eq
The last scalar term vanishes from the commutator in the equations
(\ref{vv0142}): $\dot S=[S,J^\eta(S)]$.

\subsection{Spin chains and Gaudin models}

The Lax matrix (\ref{vv7025}), (\ref{vv352})-(\ref{vv356}) allows to
define a class of integrable ${\rm gl}_N$ spin chains with the
transfer-matrix
 \beq\label{vv441}
 \begin{array}{c}
  \displaystyle{
\hat T_n(z)= L^{\eta}(\hat S^1, z-z_1)\,...\, L^{\eta}(\hat S^n,
z-z_n)\,.
 }
 \end{array}
 \eq
on $n$ sites with inhomogenuities $z_k$. The underlying quantum
algebra consists of $n$ copies of (\ref{vv007}) with the quantum
$R$-matrix (\ref{vv35222})-(\ref{vv35622}). The non-relativistic
limit gives rise to the Gaudin model defined by the Lax operator
 \beq\label{vv446}
 \begin{array}{c}
  \displaystyle{
 \hat L^G(z)=\sum\limits_{a=1}^n l(z-z_a,\hat S^a)\,.
 }
 \end{array}
 \eq
 Its Hamiltonians are computed as residues of $\tr
 \left({\hat L}^G(z)\right)^2$:
 \beq\label{vv447}
 \begin{array}{c}
  \displaystyle{
 \hat h_a=\sum\limits_{c\neq a}^n h_{a,c}\,,\ \ \
  h_{a,c}=\tr_{12}\left(r_{12}(z_a-z_c)\hat S^a_1 \hat S^c_2\right)=\tr \left(\hat S^a\,l(z_a-z_c,\hat
 S^c)\right)\,.
 }
 \end{array}
 \eq
For example, in ${\rm gl}_2$ case the classical $r$-matrix
(\ref{vv0031}) (classical limit of (\ref{vv714}))
 \beq\label{w37}
   \displaystyle{
{r}_{12}(z)=
 \left(\begin{array}{cccc}
1/z & 0 & 0 & 0\\ -z & 0 & 1/z & 0\\ -z & 1/z & 0 & 0\\ -z^3 & z & z
& 1/z
 \end{array}
 \right)
 }
 \eq
gives
 \beq\label{vv448}
 \begin{array}{c}
  \displaystyle{
  h_{a,c}=\frac{\tr(\hat S^a \hat S^c)}{z_a-z_c}
  -(z_a-z_c)\left(\hat S_{12}^a(\hat S_{11}^c-\hat S_{22}^c)+\hat S_{12}^c(\hat S_{11}^a-\hat S_{22}^a)\right)
  -(z_a-z_c)^3\,\hat S_{12}^a\hat S_{12}^c\,.
 }
 \end{array}
 \eq
In the limit (\ref{vv71400}) this formula reproduces the well-known
rational Gaudin Hamiltonians (the first term in (\ref{vv448})).

Let us also compute the quantum local Hamiltonian of the homogeneous
($z_a=0$) periodic spin chain on $n$ sites. The quantum $R$-matrix
 \beq\label{vv679}
 \begin{array}{c}
  \displaystyle{
  \tilde R^\eta(z)=z\eta R^\eta(z)
 }
 \end{array}
 \eq
with $R$ (\ref{vv714}), satisfies
 \beq\label{vv6791}
 \begin{array}{c}
  \displaystyle{
  \tilde R^\eta(0)_{12}=\eta P_{12}\,.
 }
 \end{array}
 \eq
Therefore, we can calculate the local Hamiltonian using standard
approach of \cite{Baxter,Sklyanin0}. The answer  is given by
$\sum\limits_{k=1}^n\, H_{k,k+1}$, where
$H_{k,k+1}=P_{k,k+1}\frac{d}{dz}\tilde
R^\eta_{k,k+1}(z)\left.\right|_{z=0}$:
%
%
%
%
 \beq\label{vv6497}
 \begin{array}{c}
  \displaystyle{
  H^{\hbox{\tiny{local}}}=\sum\limits_{k=1}^n\, P_{k,k+1}-\eta^2 \mathrm E_{21}^{k}\otimes(\mathrm E_{11}^{k+1}-\mathrm E_{22}^{k+1})
  -\eta^2 (\mathrm E_{11}^{k}-\mathrm E_{22}^{k})\otimes\mathrm
  E_{21}^{k+1}-\eta^4 \mathrm E_{21}^{k}\otimes\mathrm
  E_{21}^{k+1}\,,
 }
 \end{array}
 \eq
 where $\mathrm E_{ij}^{n+1}=\mathrm E_{ij}^1$.
It is a deformation of the XXX spin chain. The latter is described
by the only first term in (\ref{vv4497}):
 \beq\label{vv6498}
 \begin{array}{c}
  \displaystyle{
  H^{\hbox{\tiny{XXX}}}=\sum\limits_{k=1}^n\, P_{k,k+1}\,,\ \ \
  P_{k,k+1}=\mathrm E_{11}^{k}\otimes\mathrm
  E_{11}^{k+1}+\mathrm E_{12}^{k}\otimes\mathrm
  E_{21}^{k+1}+\mathrm E_{21}^{k}\otimes\mathrm
  E_{12}^{k+1}+\mathrm E_{22}^{k}\otimes\mathrm
  E_{22}^{k+1}\,.
 }
 \end{array}
 \eq
 The generators $\mathrm
 E_{ij}$, $(\mathrm E_{ij})_{ab}=\delta_{ia}\delta_{jb}$ are dual to $\hat S_{ji}=\tr (\mathrm E_{ij}\hat
 S)$.

The Hamiltonian of type (\ref{vv6497}) was obtained in \cite{Khor}
using different $R$-matrix (it depends on two spectral parameters)
\footnote{Another deformation of the Heisenberg chain was found in
\cite{Kulish}.}. We describe this type of models and related soliton
equations in our next publication \cite{LOZ6}.

\section{Quantum rational R-matrix}\label{quant}
\setcounter{equation}{0}

The quantum non-dynamical $R$-matrix can be found by the standard
procedure of the IRF-Vertex Correspondence starting from the
rational RS model. Here we use another approach based on
(\ref{vv707}). Applying it to (\ref{vv352})-(\ref{vv356}) we get
%
%
{\small{
 \beq\label{vv35222}
 \begin{array}{c}
  \displaystyle{
 R^\hbar_{12}(z)=\sum\limits_{k,l=1}^N\frac{\p {L}^{\hbar}(z) }{\p S_{kl}}\otimes {\mathrm E}_{lk}
 =\frac{1}{z}\sum\limits_{i,j=1}^N
 {\mathrm E}_{ij}\otimes
 \left\{\,\,\sum\limits_{\ga=0}^{\varrho(i)} z^\gamma \left(\!\begin{array}{c} \varrho(i) \\ \gamma
 \end{array}\!\right)
 {\mathrm E}_{j,\,\varrho^{-1}(\varrho(i)-\ga)}\right.
}
 \\
 \displaystyle{
  -\sum\limits_{\ga=0}^{\varrho(i)} z^{\ga+N-j+1}\,
 (-1)^{\varrho(j)+N}(N\!-\!j)\left(\!\begin{array}{c} \varrho(i) \\ \gamma
 \end{array}\!\right) \left(\!\begin{array}{c} N \\ j\!-\!1
 \end{array}\!\right) {\mathrm E}_{N,\,\varrho^{-1}(\varrho(i)-\ga)}
 }
\\
  \displaystyle{
+\!\sum\limits_{\ga=0}^{\varrho(i)} \,\sum\limits_{s=1}^{N-j}
z^{s+\ga}\, (-1)^{\varrho(j)+s+j-1} \!\left(\!\begin{array}{c} \varrho(i) \\
\gamma
 \end{array}\!\right) \left(\!\begin{array}{c} s\!+\!j\!-\!1 \\ j\!-\!1
 \end{array}\!\right)
 {\mathrm E}_{\varrho^{-1}(s+j-1),\,\varrho^{-1}(\varrho(i)-\ga)}\,-
 }
 \end{array}
 \eq
 \beq\label{vv35322}
 \begin{array}{c}
 \displaystyle{
  -N\sum\limits_{s=1}^{N-j}\, \sum\limits_{b=0}^{\varrho(i)}
  (-1)^{\varrho(j)\!+s+j-1}\, z^s\, (z+\hbar)^b \left(\!\begin{array}{c} s\!+\!j\!-\!2 \\ j\!-\!1
 \end{array}\!\right) \left(\!\begin{array}{c} \varrho(i) \\ b
 \end{array}\!\right)\times
  }
   \\
 \displaystyle{\left[
 \delta_{\varrho(i)-j-s-b+1\leq\, 0} \sum\limits_{c=0}^{N\!-s-j+1}\ \sum\limits_{p=0}^{\varrho(i)\!-b+c}
   (-\hbar)^p\left(\!\begin{array}{c} \varrho(i)\!-\!b\!+\!c \\ p
 \end{array}\!\right)  {\mathrm E}_{\varrho^{-1}(s+j+c-1),\,\varrho^{-1}(\varrho(i)-b-p+c)}
  \right.}
     \\
 \displaystyle{\left.
\left. -\delta_{\varrho(i)-j-s-b+1>\, 0} \sum\limits_{c=0}^{s+j-2}\
\sum\limits_{p=0}^{\varrho(i)\!-b-c-1}
   (-\hbar)^p\left(\!\begin{array}{c} \varrho(i)\!-\!b\!-\!c\!-\!1 \\ p
 \end{array}\!\right)  {\mathrm E}_{\varrho^{-1}(s+j-c-2),\,\varrho^{-1}(\varrho(i)-b-p-c-1)}
  \right]\,\right\}}
 \end{array}
 \eq
 \beq\label{vv35422}
 \begin{array}{c}
  \displaystyle{
+\frac{1}{z}\left[zN\,{\mathrm E}_{NN}-\sum\limits_{j=1}^N
z^{N-j+2}(-1)^{\varrho(j)+N} (N-j)N \left(\!\begin{array}{c} N \\
j-1 \end{array}\!\right) {\mathrm E}_{Nj}\right.
  }
  \\
   \displaystyle{
  -N\sum\limits_{i,j=1}^N\,\sum\limits_{s=1}^{N-j}\, \sum\limits_{b=0}^{\varrho(i)}
  (-1)^{\varrho(j)\!+s+j-1}\, z^s\, (z+\hbar)^b \left(\!\begin{array}{c} s\!+\!j\!-\!2 \\ j\!-\!1
 \end{array}\!\right) \left(\!\begin{array}{c} \varrho(i) \\ b
 \end{array}\!\right)\delta_{p\leq 1}\,\delta_{\varrho(i)-b-p-j-s+2\,,\,0}\,\times
  }
     \\
 \displaystyle{
 \left.\sum\limits_{p=0}^{\varrho(i)-b+N-s-j+1}
   (-\hbar)^p\left(\!\begin{array}{c} \varrho(i)-b+N\!-s-j+1 \\ p
 \end{array}\!\right)\, {\mathrm E}_{ij}\,  \right] \otimes
 }
 \end{array}
 \eq
 \beq\label{vv35622}
 \begin{array}{c}
  \displaystyle{
\otimes\left(-\frac{1}{N(-\hbar)}\sum\limits_{j=1}^N {\mathrm
E}_{jj}-\frac{1}{N^2}\sum\limits_{j=1}^N\Big[\delta_{\varrho(j)\geq
1} \,\varrho(j)\, {\mathrm
E}_{\varrho^{-1}(j),\,\varrho^{-1}(\varrho(j)-1)}+(-1)^{\varrho(j)+j}\,
j\,\, {\mathrm E}_{\varrho^{-1}(j),\,j}\right.
  }
  \\
  \displaystyle{\left.
  -N\sum\limits_{b=0}^{\varrho(j)} (-1)^{\varrho(j)\!+\!j+b}\, \left(\!\begin{array}{c} \varrho(j) \\ b
 \end{array}\!\right)
 \sum\limits_{c=0}^{N-j}\ \sum\limits_{p=0}^{\varrho(j)\!-b+c}
   (-\hbar)^{p+b}\left(\!\begin{array}{c} \varrho(j)\!-\!b\!+\!c \\ p
 \end{array}\!\right)
 {\mathrm E}_{\varrho^{-1}(j+c),\,\varrho^{-1}(\varrho(j)-b-p+c)}\Big]\right)
  }
 \end{array}
 \eq
 }}
As in (\ref{vv352})-(\ref{vv356}) we imply that the values of
indices corresponding to undefined argument value $N-1$ of
$\varrho^{-1}$ function are skipped in summations.
Notice that the obtained $R$-matrix is unitary (\ref{vv00271}) with
$f^\hbar(z)=\frac{1}{\hbar^2}-\frac{1}{z^2}$:
 \beq\label{vv3570}
 \begin{array}{c}
  \displaystyle{
 R^\hbar_{12}(z)R^\hbar_{21}(-z)=\left(\frac{1}{\hbar^2}-\frac{1}{z^2}\right)1\otimes
 1\,.
 }
 \end{array}
 \eq
By redefinition
 \beq\label{vv35705}
 \begin{array}{c}
  \displaystyle{
 R^\hbar_{12}(z,\epsilon)=\epsilon\,R^{\epsilon\,\hbar}_{12}(z\epsilon)
 }
 \end{array}
 \eq
we can treat the answer as deformation of the standard XXX
$R$-matrix. Indeed, one can verify that
 \beq\label{vv35706}
 \begin{array}{c}
  \displaystyle{
 \lim\limits_{\epsilon\to 0} R^\hbar_{12}(z,\epsilon)=\hbar^{-1}
 1\otimes 1+z^{-1} P_{12}\,.
 }
 \end{array}
 \eq
 In the classical limit we get the rational skew-symmetric non-dynamical
 $r$-matrix from \cite{AASZ}\footnote{Expression (\ref{v42002}) differs from the one given in \cite{AASZ} by common factor $N$ and scalar
 term $1\otimes 1/Nz$.}:
 {\small{
  \beq\label{v42002}
 {r}^{\hbox{\tiny{top}}}(z)=\frac{1}{Nz}1\otimes 1+\frac{1}{z}\sum\limits_{i,j=1}^N E_{ij}\,\otimes
 \eq
 $$
 \begin{array}{c}
 \displaystyle{
 \Big[\sum\limits_{\ga=0}^{\varrho(i)} z^\gamma \left(\!\begin{array}{c} \varrho(i) \\ \gamma
 \end{array}\!\right)
 E_{\varrho^{-1}(\varrho(i)-\ga),j}-\sum\limits_{\ga=0}^{\varrho(i)} z^{\ga+N-j+1}\,
 (-1)^{\varrho(j)+N}(N\!-\!j)\left(\!\begin{array}{c} \varrho(i) \\ \gamma
 \end{array}\!\right) \left(\!\begin{array}{c} N \\ j\!-\!1
 \end{array}\!\right) E_{\varrho^{-1}(\varrho(i)-\ga),\,N}
 }
 \\
  \displaystyle{
-N\sum\limits_{\ga=0}^{\varrho(i)} \sum\limits_{s=1}^{N-j}
\delta_{\varrho(i)-j+1\leq s+\ga} z^{s+\ga}\,
(-1)^{\varrho(j)+s+j-1}\left(\!\begin{array}{c} \varrho(i) \\
\gamma
 \end{array}\!\right) \left(\!\begin{array}{c} s\!+\!j\!-\!2 \\ j\!-\!1
 \end{array}\!\right)\sum\limits_{c=0}^{N-s-j+1} E_{\varrho^{-1}(\varrho(i)\!-\!\ga\!+\!c),\,\varrho^{-1}(s\!+\!j\!+\!c\!-\!1)}
 }
  \end{array}
 $$
 $$
 \begin{array}{c}
  \displaystyle{
-N\sum\limits_{\ga=0}^{\varrho(i)} \sum\limits_{s=1}^{N-j}
\delta_{\varrho(i)-j+1> s+\ga} z^{s+\ga}\,
(-1)^{\varrho(j)+s+j-2}\left(\!\begin{array}{c} \varrho(i) \\
\gamma
 \end{array}\!\right) \left(\!\begin{array}{c} s\!+\!j\!-\!2 \\ j\!-\!1
 \end{array}\!\right)\sum\limits_{c=0}^{s+j-2} E_{\varrho^{-1}(\varrho(i)\!-\!\ga\!-\!c\!-1),\,\varrho^{-1}(s\!+\!j\!-\!c\!-2)}
 }
 \\
\displaystyle{ +\!\sum\limits_{\ga=0}^{\varrho(i)}
\sum\limits_{s=1}^{N-j}
z^{s+\ga}\, (-1)^{\varrho(j)+s+j-1} \left(\!\begin{array}{c} \varrho(i) \\
\gamma
 \end{array}\!\right) \left(\!\begin{array}{c} \!s\!+\!j\!-\!1\! \\ \!j\!-\!1\!
 \end{array}\!\right) E_{\varrho^{-1}(\varrho(i)-\ga),\,\varrho^{-1}(s+j-1)}\!-\!\frac{\delta_{i,j}}{N}
\sum\limits_{k=1}^N\sum\limits_{c=0}^{N\!-\!k\!-2}E_{k+c,\,k+c+1}\Big]
 }
 \end{array}
 $$
 }}
\vskip3mm

\noindent {\bf Example: 11-vertex R-matrix.}  In $N=2$ case we
obtain the 11-vertex R-matrix \cite{Cherednik}:
 \beq\label{vv320}
 \begin{array}{c}
  \displaystyle{R^\hbar(z)=
 \left( \begin{array}{cccc} {\hbar}^{-1}+{z}^{-1}&0&0&0
\\\noalign{\medskip}-\hbar-z&{\hbar}^{-1}&{z}^{-1}&0\\\noalign{\medskip}
-\hbar-z&{z}^{-1}&{\hbar}^{-1}&0\\\noalign{\medskip}-{\hbar}^{3}-2\,z{\hbar}^{2}-2\,\hbar
\,{z}^{2}-{z}^{3}&\hbar+z&\hbar+z&{\hbar}^{-1}+{z}^{-1}
\end{array} \right)
  }
 \end{array}
 \eq

\vskip3mm

\noindent {\bf Example: rational ${\rm gl}_3$ $R$-matrix.}
%
%
%
%
%
In ${\rm gl}_3$ case (\ref{vv35222})-(\ref{vv35622}) gives the
following $9\times 9$ quantum $R$-matrix:
 \beq\label{vv321}
 \begin{array}{c}
  \displaystyle{
  R^\hbar(z)=
  }
 \end{array}
 \eq
{\footnotesize{
 $$
 \left( \begin{array}{ccc} {\hbar}^{-1}+{z}^{-1}&0&0
\\\noalign{\medskip}1&{\hbar}^{-1}&0\\\noalign{\medskip}2\,{\hbar}^{2}+3
\,z\hbar+2\,{z}^{2}&-3\,\hbar-3\,z&{\hbar}^{-1}\\\noalign{\medskip}-1&{z}
^{-1}&0\\\noalign{\medskip}2\,\hbar+2\,z&0&0\\\noalign{\medskip}2\,{z}^
{3}+3\,z{\hbar}^{2}+2\,{\hbar}^{3}+3\,{z}^{2}\hbar&-3\,{\hbar}^{2}-3\,z
\hbar-{z}^{2}&1\\\noalign{\medskip}-2\,{\hbar}^{2}-3\,z\hbar-2\,{z}^{2}&-
3\,\hbar-3\,z&{z}^{-1}\\\noalign{\medskip}2\,{z}^{3}+3\,z{\hbar}^{2}+2\,
{\hbar}^{3}+3\,{z}^{2}\hbar&3\,{z}^{2}+3\,z\hbar+{\hbar}^{2}&-1
\\\noalign{\medskip}2\,{\hbar}^{5}+3\,{z}^{4}\hbar+3\,{z}^{2}{\hbar}^{3}+
2\,{z}^{5}+3\,z{\hbar}^{4}+3\,{z}^{3}{\hbar}^{2}&3\,{z}^{4}-3\,{\hbar}^{4
}-3\,z{\hbar}^{3}+3\,{z}^{3}\hbar&-{z}^{2}+{\hbar}^{2}\end{array}
 \right.
 $$

 $$
 \left. \begin{array}{cccccc} 0&0&0&0&0&0\\\noalign{\medskip}{z}^{-1}
&0&0&0&0&0\\\noalign{\medskip}-3\,\hbar-3\,z&3&0&{z}^{-1}&0&0
\\\noalign{\medskip}{\hbar}^{-1}&0&0&0&0&0\\\noalign{\medskip}0&{\hbar}^
{-1}+{z}^{-1}&0&0&0&0\\\noalign{\medskip}-3\,z\hbar-3\,{z}^{2}-{\hbar}^{
2}&0&{\hbar}^{-1}&1&{z}^{-1}&0\\\noalign{\medskip}-3\,\hbar-3\,z&-3&0&{
\hbar}^{-1}&0&0\\\noalign{\medskip}{z}^{2}+3\,{\hbar}^{2}+3\,z\hbar&0&{z}
^{-1}&-1&{\hbar}^{-1}&0\\\noalign{\medskip}3\,z{\hbar}^{3}+3\,{\hbar}^{4}
-3\,{z}^{3}\hbar-3\,{z}^{4}&-6\,{\hbar}^{3}-6\,{z}^{3}-9\,z{\hbar}^{2}-9
\,{z}^{2}\hbar&3\,z+3\,\hbar&-{\hbar}^{2}+{z}^{2}&3\,z+3\,\hbar&{\hbar}^{-1
}+{z}^{-1}\end {array}\right)
 $$
}}
Plugging it into (\ref{vv7142}) one gets the ${\rm gl}_3$
relativistic top. In (\ref{vv380}) case it
gauge equivalent to 3-body rational RS model via (\ref{vv515}).

\noindent Remark: Presumably, the $R$-matrix
(\ref{vv35222})-(\ref{vv35622}) can be obtained by special limiting
procedure from the Belavin's elliptic $R$-matrix. Such an algorithm
was described in \cite{Smirnov}. Computer calculations gave the same
answer for $N=2,3$ (as in our examples (\ref{vv320}),
(\ref{vv321})).
There is another approach to the rational $R$-matrices based on the
deformed Yangians \cite{Kulish}. The $R$-matrices  in this
description depend on two spectral parameters. It seems likely that
some relation to our $R$-matrix may exist. For example, the twisted
${\rm gl}_2$ case considered in \cite{Khor} leads to the same local
spin chain Hamiltonian. In the same time the deformation discussed
in \cite{Kulish} differs from ours. Let us also notice that an
algorithm for computing ${\rm gl}_{N}$ quantum $R$-matrices (related
to \cite{Kulish} description) was proposed and studied in
\cite{Burban}.

\section{Belavin's $R$-matrix and elliptic models}\label{ell}
\setcounter{equation}{0}

\subsection{Sklyanin algebra and relativistic elliptic tops}


\noindent { Belavin's   elliptic ${\rm gl}_N$ $R$-matrix}
\cite{Belavin}
 \beq\label{vv005}
 \begin{array}{c}
  \displaystyle{
 R^\hbar_{12}(z)=\sum\limits_{a\in\,{\mathbb Z}_N\times{\mathbb Z}_N} \vf_a^\hbar(z)\, T_a\otimes
 T_{-a}
 }
 \end{array}
 \eq
is the central object of the section as well as related quantum
$L$-operator:
 \beq\label{vv006}
 \begin{array}{c}
  \displaystyle{
 \hat L^\eta(z)\stackrel{(\ref{vv0025})}{=}\tr_2 \left(R^{\,\eta}_{12}(z)\hat S_2\right)=
 \sum\limits_{a\in\,{\mathbb Z}_N\times{\mathbb Z}_N} \vf_a^\eta(z)\,
 T_a\,\hat S_a\,.
 }
 \end{array}
 \eq
In (\ref{vv005}) and (\ref{vv006}) the basis of ${\rm gl}_N$ and
corresponding functions are chosen as\footnote{See for example
review \cite{SZ}.}
 \beq\label{vv401}
 \begin{array}{c}
  \displaystyle{
 T_a=T_{a_1 a_2}=\exp\left(\frac{\pi\imath}{N}\,a_1
 a_2\right)Q^{a_1}\Lambda^{a_2}\,,\ \ \ Q_{kl}=\delta_{kl}\exp(\frac{2\pi
 i}{N}k)\,,\ \ \ \Lambda_{kl}=\delta_{k-l+1=0\,{\hbox{\tiny{mod}}}
 N}\,,
 }
 \end{array}
 \eq
 \beq\label{vv402}
 \begin{array}{c}
  \displaystyle{
\vf_a^\eta(z)=\exp(2\pi\imath z\p_\tau\om_a)\phi(z,\om_a+\eta)\,,\ \
\phi(z,u)=\frac {\vth'(0)\vth(u+z)} {\vth(z)\vth(u)}\,,\ \
\om_a=\frac{a_1+a_2\tau}{N}\,,
 }
 \end{array}
 \eq
where $\vth(z)$ is the odd theta function and $a_1\,,a_2\in {\mathbb
Z}_N$.

The Sklyanin algebra (\ref{vv007}) is defined by the local behavior
of $\vf_a^\eta(z)$ near $z=0$:
 \beq\label{vv403}
 \begin{array}{c}
  \displaystyle{
\phi(z,u)=\left(\frac{1}{z}+E_1(u)+\frac{z}{2}(E_1^2(u)-\wp(u))+\ldots\right)\,,\
\ E_1(z)=\p_z \log\vth(z)\,.
 }
 \end{array}
 \eq
Then
 \beq\label{vv404}
 \begin{array}{c}
  \displaystyle{
R^{\hbar,(0)}_{12}(z)=\sum\limits_{a\in\,{\mathbb Z}_N\times{\mathbb
Z}_N} E_1(\om_a+\hbar)\, T_a\otimes
 T_{-a}\,,
 }
 \end{array}
 \eq
%
and
 \beq\label{vv405}
 \begin{array}{c}
  \displaystyle{
 \hat L^{\eta,(0)}(z)=
 \sum\limits_{a\in\,{\mathbb Z}_N\times{\mathbb Z}_N} (E_1(\om_a+\eta)+2\pi\imath\p_\tau\om_a)\,
 T_a\,\hat S_a\,.
 }
 \end{array}
 \eq
The last terms ($2\pi\imath\p_\tau\om_a$) are canceled out in the
final answers.

\noindent {\bf Sklyanin algebra ${\mathcal
A}_{\hbar,\eta}^{\hbox{\tiny{Skl}}}$} for $\hbar,\eta\neq 0$ in
components $\hat S_a=\tr(\hat S\, T_{-a})$ can be derived from
either (\ref{vv007}) together with (\ref{vv404}), (\ref{vv405}) or
directly -- by plugging (\ref{vv005}) and (\ref{vv006}) into
exchange relations (\ref{vv001}). The latter way requires identity
 $$
 \vf_{a-c}^\eta(z)\vf_{b+c}^\eta(w)\vf_c^\hbar(z-w)-\vf_{b+c}^\eta(z)\vf_{a-c}^\eta(w)\vf_{a-b-c}^\hbar(z-w)=
 \bff(a,b,c|\tau,\hbar,\eta)
  \,\vf_a^{\eta+\hbar}(z)\vf_b^{\eta-\hbar}(w)\,,
 $$
 where
 \beq\label{vv4059}
 \begin{array}{c}
   \displaystyle{
 \bff(a,b,c|\tau,\hbar,\eta)=E_1(\om_c+\hbar)-E_1(\om_{a-b-c}+\hbar)+E_1(\om_{a-c}+\eta)-E_1(\om_{b+c}+\eta)\,.
 }
 \end{array}
 \eq
The quadratic relations ${\mathcal
A}_{\tau,\hbar,\eta}^{\hbox{\tiny{Skl}}}$ in the $T_a\otimes T_b$
component of (\ref{vv001}) read as follows:
 \beq\label{vv40591}
 \begin{array}{c}
  \displaystyle{
 \sum\limits_{c\in\,{\mathbb Z}_N\times{\mathbb Z}_N}\,
\bff(a,b,c|\tau,\hbar,\eta) \left( \hat S_{a-c}\,\hat
S_{b+c}\,\kappa_{c,a-b}-\hat S_{b+c}\,\hat S_{a-c}\,\kappa_{a-b,c}
 \right)=0\,,
 }
 \end{array}
 \eq
where $\kappa_{a,b}=\exp\frac{\pi\imath}{N}(a_2 b_1-a_1 b_2)$ comes
from $T_a\,T_b=\kappa_{a,b}\,T_{a+b}$.
Notice that in the case $\eta=\hbar$ and $b=0$ one should consider
the limit $\eta\rightarrow \hbar$ in (\ref{vv4059}) which gives
 \beq\label{vv40592}
 \begin{array}{c}
  \displaystyle{
 \vf_{a-c}^\hbar(z)\vf_{b+c}^\hbar(w)\vf_c^\hbar(z-w)-\vf_{b+c}^\hbar(z)\vf_{a-c}^\hbar(w)\vf_{a-b-c}^\hbar(z-w)=
 \bff(a,0,c|\tau,\hbar,\hbar)\,\vf_a^{2\hbar}(z)\,,
 }
 \\ \ \\
   \displaystyle{
 \bff(a,0,c|\tau,\hbar,\hbar)=E_2(\om_c+\hbar)-E_2(\om_{a}-\om_c+\hbar)\,.
 }
 \end{array}
 \eq
The case $\eta=\hbar$ and $N=2$ in (\ref{vv40591}) gives rise to the
Sklyanin algebra in its original form \cite{Sklyanin}.

It follows from (\ref{vv4059}) that the "structure constant"
$\bff(a,b,c|\tau\hbar,\eta)$ is double periodic with respect to the
shifts
$$
\hbar\to\hbar+\mZ+\mZ\tau\,,~~\eta\to\eta+\mZ+\mZ\tau\,.
$$
Therefore, we can consider the pair $(\hbar,\eta)$ as points on two
elliptic curves $\Si_\tau=\mC/(\mZ+\tau\mZ)$.
Consider the upper half-plane $\cH^+\subset\mC$. The moduli space
$\gM$ of elliptic curves is the result of the action of $\SLZ$ on
$\cH^+$ by the M\"obius transform
$$
\gM=\cH^+/\SLZ\,,~~\left(\tau\to
\frac{\al\tau+\be}{\ga\tau+\delta}\right)\,.
$$
The modular transformations acts on the theta-function as 
$$
\vth\left(\frac{v}{\al+\be\tau}|\frac{\ga+\de\tau}{\al+\be\tau}\right)=\zeta
(\al+\be\tau)^\oh\exp\left(\frac{\imath\pi\al
v^2}{\al+\be\tau}\right)\vth(v|\tau)\,,
$$
where $\zeta^8=1$. Then we can find that
$\bff(a,b,c|\tau,\hbar,\eta)$ is modular invariant, and
 therefore,
it is a well defined function on  $\gM$.
In this way the universal bundle \beq\label{bms} \cE_\tau=
\begin{array}{c}
  \Si_\tau\times\Si_\tau \\
  \downarrow \\
  \gM
\end{array}
\eq plays the role of the moduli space of the algebra ${\mathcal
A}_{\tau,\hbar,\eta}^{\hbox{\tiny{Skl}}}$.
Then the moduli space is \beq\label{msp} Mod\,({\mathcal
A}_{\tau,\hbar,\eta}^{\hbox{\tiny{Skl}}})=\cE_\tau\,. \eq

\noindent {\bf Relation between ${\mathcal A}_{\eta\neq
0}^{\hbox{\tiny{Skl}}}$ and ${\mathcal
A}_{\eta=0}^{\hbox{\tiny{Skl}}}$.} Following \cite{CLOZ}\footnote{In
that paper $\eta=\hbar$ was considered.} let us clarify the relation
between $\eta$-dependent and $\eta$-indepen\-dent $L$-operators
discussed above (\ref{vv0314}). Two descriptions are distinct from
each other by quasiperiodic boundary conditions on the lattice
$\mathbb C/\mathbb Z+\mathbb Z\tau$:
 \beq\label{vv4055}
 \begin{array}{c}
  \displaystyle{
 L^\eta(z+1)=Q^{-1} L^\eta(z) Q\,,\ \ \  L^\eta(z+\tau)=\exp(-2\pi\imath\,\eta)\Lambda^{-1}
 L^\eta(z) \Lambda\,,
 }
 \end{array}
 \eq
 \beq\label{vv4056}
 \begin{array}{c}
  \displaystyle{
 \mathrm L(z+1)=Q^{-1} \mathrm L(z) Q\,,\ \ \  \mathrm L(z+\tau)=\Lambda^{-1}
 \mathrm L(z) \Lambda\,,
 }
 \end{array}
 \eq
where $Q$ and $\Lambda$ are from  (\ref{vv401}). Notice that
$\mathrm L(z)$ (\ref{vv4056}) is a section of $\hbox{End}V$-bundle,
where $V$ is the holomorphic vector bundle with the transition
functions $Q$ and
$\ti\Lambda(z)=\exp(2\pi\imath(\frac{z}{N}+\frac{\tau}{2N}))\Lambda$.
In the same time $L^\eta(z)$ should be considered as a map between
$V$ and $V'$, where $V'$ is defined by the transition functions $Q$
and $\Lambda'(z)=\exp(2\pi\imath \eta)\Lambda(z)$.

Set $\eta_0=-\eta$ and $g^\eta(z)=1/\phi(z-\eta,\eta)$ in
(\ref{vv0314}). Then
 \beq\label{vv406}
 \begin{array}{c}
  \displaystyle{
 \frac{1}{\phi(z-\eta,\eta)}\hat L^{\eta}(z-\eta,\vf(\hat S))=T_0\hat S_0+\sum\limits_{a\neq 0} \vf_a^0(z)\,
 T_a\,\hat S_a\,,
 }
 \end{array}
 \eq
where
 \beq\label{vv407}
 \begin{array}{c}
  \displaystyle{
 \vf(\hat S)=T_0\hat S_0+\sum\limits_{a\neq 0} \vf_a^0(\eta)\,
 T_a\,\hat S_a\,,
 }
 \end{array}
 \eq
i.e. the $\eta$-independent description is given by
 \beq\label{vv4073}
 \begin{array}{c}
  \displaystyle{
 \mathrm L(z,\hat S)=T_0\hat S_0+\sum\limits_{a\neq 0} \vf_a^0(z) T_a\,\hat S_a=T_0\hat S_0+\tr_2(r_{12}(z)\hat S)\,
\,,
 }
 \end{array}
 \eq
 The underlying identity is very simple:
 \beq\label{vv408}
 \begin{array}{c}
  \displaystyle{
\vf^\eta_a(z-\eta)/\vf^\eta_0(z-\eta)=\vf_a^0(z)/\vf^0_a(\eta)\,.
 }
 \end{array}
 \eq
%
The sum over $a\neq 0$ corresponds to ${\rm sl}_N$ part, i.e. $\hat
S$ can be replaced with $\hat{\bar S}$ in the r.h.s. of
(\ref{vv406}) and (\ref{vv407}) as in (\ref{vv03117}). The classical
$r$-matrix (\ref{vv003}) emerging in the $\eta$-independent form
(\ref{vv03117}) is the Belavin-Drinfeld ${\rm sl}_N$ $r$-matrix
\cite{BD}:
 \beq\label{vv409}
 \begin{array}{c}
  \displaystyle{
 r_{12}(z)=\sum\limits_{a\neq 0} \vf_a^0(z)\, T_a\otimes
 T_{-a}\,.
 }
 \end{array}
 \eq
The scalar term $T_0\otimes T_0\, E_1(z)$ is not important here.

\noindent {\bf Relativistic top} appears in the quasi-classical
limit. The Lax matrix (\ref{vv0027})\footnote{In (\ref{vv0027})
$\stackrel{}{\mathcal
 J^\eta}_{a,b}(z)=\delta_{a+b}\,\vf_a^\eta(z)$.}
 \beq\label{vv410}
 \begin{array}{c}
  \displaystyle{
{L^\eta}(z)=\tr_2 \left(R^{\,\eta}_{12}(z)
S_2\right)=\sum\limits_{a\in\,{\mathbb Z}_N\times{\mathbb Z}_N}
\vf_a^\eta(z)\,
 T_a\, S_a\,.
 }
 \end{array}
 \eq
together with $M$-matrix (\ref{vv016})
 \beq\label{vv411}
 \begin{array}{c}
  \displaystyle{
M(z)=-\tr_2\left(r_{12}(z)S_2\right)\stackrel{(\ref{vv409})}{=}\,-\sum\limits_{a\neq
0} \vf_a^0(z)\, T_a\,S_a
 }
 \end{array}
 \eq
provides equations of motion (\ref{vv0142}) generated by the
Hamiltonian $H=S_0$:
  \beq\label{vv412}
  \begin{array}{c}
  \displaystyle{
\p_{t_0}\,S=\{S_0,S\}=[S,J^\eta(S)]\,,
 }
 \end{array}
 \eq
where the inverse inertia tensor $J^\eta$ (\ref{vv0143}):
   \beq\label{vv413}
  \begin{array}{c}
  \displaystyle{
 J^\eta(S)=\tr_2\left(\left(R_{12}^{\eta,(0)}-r_{12}^{(0)}\right)S_2\right)=T_0 S_0 E_1(\eta)+
 \sum\limits_{a\neq 0}T_aS_a(E_1(\om_a+\eta)-E_1(\om_a))=
   }
   \\ \ \\
   \displaystyle{
 =S\, E_1(\eta)+
 \sum\limits_{a\neq 0}T_aS_a(E_1(\om_a+\eta)-E_1(\om_a)-E_1(\eta))\,.
   }
   \end{array}
  \eq
The first (scalar) term $T_0 S_0 E_1(\eta)$ in the upper line  of
(\ref{vv413})  vanishes from the commutator in (\ref{vv412}) as well
as the first term $S E_1(\eta)$ in the lower line. To verify
(\ref{vv412}), (\ref{vv413}) one needs the following identity:
   \beq\label{vv414}
  \begin{array}{c}
  \displaystyle{
 \vf_a^\eta(z)\,\vf_b^0(z)=\vf^\eta_{a+b}(z)\,\left(E_1(z)+E_1(\om_a+\eta)+E_1(\om_b)-E_1(z+\om_{a+b}+\eta)\right)\,.
   }
   \end{array}
  \eq

\noindent {\bf Non-relativistic limit} $\eta\rightarrow 0$ coincides
with $\eta$-independent description at the level of equations of
motion because the Lax matrices (\ref{vv020}) and (\ref{vv0311}) are
the same up to the scalar term $S_0\,1$. This is due to existence of
bihamiltonian structure, i.e. any linear combination of the linear
and quadratic Poisson brackets are again some Poisson bracket (see
\cite{KLO} for details).

The equations of motion (\ref{vv412}) keep the same form in the
limit with
   \beq\label{vv415}
  \begin{array}{c}
  \displaystyle{
 \mathrm J(S)=\lim\limits_{\eta\rightarrow
 0}\eta^{-1}\,J^\eta(\bar S)=-\sum\limits_{a\neq
 0}T_aS_a\,E_2(\om_a)\,,\ \ \ E_2(z)=-\p^2_z\log\vth(z)\,,
   }
   \end{array}
  \eq
  where $\bar S$ is the ${\rm sl}_N$ part of $S$.
The equations of motion $\dot S=[S,\mathrm J(S)]$ are generated by
the Lax pair $l(z)=\sum\limits_{a\neq 0} \vf^0_a(z)\, T_a\,S_a$ and
the $M$-matrix (\ref{vv039}):
   \beq\label{vv416}
  \begin{array}{c}
  \displaystyle{
\mathcal M(z)=\sum\limits_{a\neq 0} f_a(z)\, T_a\,S_a\,,\ \
f_a(z)=\p_\eta\vf_a^\eta(z)\left.\right|_{\eta=0}\,.
   }
   \end{array}
  \eq
  The underlying elliptic function identity is very well known
  \cite{Krich1}:
   \beq\label{vv4417}
  \begin{array}{c}
  \displaystyle{
\vf^0_a(z)f_b(z)-\vf^0_b(z)f_a(z)=\vf^0_{a+b}(z)(E_2(\om_a)-E_2(\om_b))=\vf^0_{a+b}(z)(\wp(\om_a)-\wp(\om_b))\,.
   }
   \end{array}
  \eq
 This model was introduced in \cite{LOZ}. Its phase space is the coadjoint orbit of Lie group ${\rm
 GL}_N$. When dimension of the orbit is minimal ($2N-2$) the Lax
 matrix is gauge equivalent to the one of elliptic Calogero-Moser model.
 More  detailed description can be found in
  \cite{SZ}. Higher rank Sklyanin algebras in the context of integrable systems were also discussed in
  \cite{BDOZ,KLO,CLOZ}.


\subsection{Poincar\'e invariance}\label{Poincare}

The Poincar\'e Lie algebra for the relativistic integrable systems
\cite{Ruijs1}
 is defined as
\beq\label{pa}
\begin{array}{l}
  \{\cH,\cP\}=0\,, \ \ \ \
  \{\cB,\cH\}=\cP\,,\ \ \ \
  \{\cB,\cP\}=\cH\,.
\end{array}
\eq
 The RS models can be obtained by symplectic (or the Poisson) reduction
 procedures from the cotangent bundles to a certain loop groups \cite{GN}\footnote{See also example of the relativistic Toda
 systems \cite{Ruijs3} in
 \cite{FockMars}}.
 For the top-like (elliptic) Lax operators (\ref{vv7115})
 a similar procedure was suggested in \cite{BDOZ} and \cite{CLOZ}.
 In all the description the Lax matrix appears through reduction from
 the group element. It satisfies some moment map constraint
 generated by the symmetries of the (co)adjoint action.

Let us show that the mentioned above reductions provide naturally
the Poincar\'e Lie algebra (\ref{pa}). As a preliminary, consider
the finite-dimensional case, i.e. the cotangent bundle $T^*G$ to
$G={\rm GL}_N$ Lie group. Let $g\in G$ and $A\in{\rm gl}_N^*$. The
symplectic structure on $T^*G$ is equal to
   \beq\label{vv580}
  \begin{array}{c}
  \displaystyle{
\om=d\,\tr\left(Ag^{-1}dg\right)\,.
   }
   \end{array}
  \eq
The corresponding Poisson brackets are of the form:
   \beq\label{vv581}
  \begin{array}{c}
  \displaystyle{
\{A_1,A_2\}=[P_{12},A_2]\,, \ \ \{g_1,g_2\}=0\,,
   }
   \end{array}
  \eq
   \beq\label{vv582}
  \begin{array}{c}
  \displaystyle{
\{A_1,g_2\}=g_1P_{12}\,,
   }
   \end{array}
  \eq
Taking $\tr_1$ of (\ref{vv582}) we get
   \beq\label{vv583}
  \begin{array}{c}
  \displaystyle{
\{\tr A,g\}=g\,.
   }
   \end{array}
  \eq
Therefore, $\{\tr A,g^{-1}\}=-g^{-1}$, and we have the following
identification with (\ref{pa}):
   \beq\label{vv584}
  \begin{array}{c}
  \displaystyle{
\cH:=\tr\left(g+g^{-1}\right),\ \ \ \
\cP:=\tr\left(g-g^{-1}\right),\ \ \ \ \cB:=\tr A\,.
   }
   \end{array}
  \eq
Notice that the variable dual to the boost $\cB$ is $\log(\det
g)^{1/N}$. Indeed, it follows from  (\ref{vv583}) that
   \beq\label{vv585}
  \begin{array}{c}
  \displaystyle{
\{\tr A,\det g\}=N\det g\,.
   }
   \end{array}
  \eq
After reduction by the action of the gauge group
   \beq\label{vv586}
  \begin{array}{c}
  \displaystyle{
g\to f^{-1}g f\,, \ \ \ A\to f^{-1}A f\,,\ \ \ f=f(A)
   }
   \end{array}
  \eq
the trivial Poisson brackets $\{g_1,g_2\}=0$ become those quadratic
of the $r$-matrix form (\ref{vv7114}) \cite{BDOZ}. The quantities
$\cH$, $\cP$ and $\cB$ from (\ref{vv584}) are gauge invariants. The
Poisson brackets (\ref{pa}) of the Poincar\'e Lie algebra keep the
same form after reduction.

Similar reasoning is valid in more complicated cases. In \cite{CLOZ}
the classical Lax operator $\mathrm L(z,S)$ (\ref{vv4073}) (where
$\hat S\to S$) was derived starting from the affine space over the
cotangent bundle $T^*LL(G)$ to the two-loop group $LL(G)$, $G={\rm
GL}_N$.
Here we deal with the holomorphic bundle on the elliptic curve
$\Sigma_\tau$ (with moduli $\tau$) given by transition functions $Q$
and $\ti\Lambda=\exp
(2\pi\imath\left(-\frac{z}{N}-\frac{\tau}{2N}\right))\Lambda$ from
(\ref{vv401}).
$\bA$ -- is the component of the $d_{\bA}$ connection on
$\Sigma_\tau$.
 In the Dolbeault description the holomorphic
structure of the  vector bundle  is defined by the operator
   \beq\label{qa}
  \begin{array}{c}
  \displaystyle{
d_{\bA}=\bp+\bA\,:\,\Om^{(0,0)}(\Si_\tau,\gln)\to\Om^{(0,1)}(\Si_\tau,\gln)
   }
   \\ \ \\
  \displaystyle{
\bA(z+1,\bz+1)=Q\bA(z,\bz)Q^{-1}\,,~~\bA(z+\tau,\bz+\bar\tau)=\La
\bA(z,\bz)\La^{-1}\,.
   }
   \end{array}
  \eq
A section $\xi(z,\bz)$ is holomorphic if $d_{\bA}\xi(z,\bz)=0$.
  Two holomorphic structures $d_{\bA}$
and $d_{\bA'}$ are equivalent if they are related by the gauge
transformation of the gauge group $\cG$
 \beq\label{gta}
 \bA\to f^{-1}\bA f+f^{-1}\bp f\,.
  \eq
 The
quotient of the space of generic connections $\cA=\{d_{\bA}\}$
\footnote{We don't discuss here the stability of the bundles,
because we need only an open subset of $\cA$.} by the gauge group
action is the moduli space $Bun(\Si_\tau,\GLN)=\cA/\cG$
 of holomorphic bundles.
The initial phase space ${\rm P}_{\Si_\tau}$ is the Poisson algebra
of holomorphic functionals on $\cR$. The Poisson brackets are
similar to (\ref{vv581}), (\ref{vv582}) (see \cite{CLOZ}). The gauge
transformations (\ref{gta}) along with
 \beq\label{gla}
g\to f^{-1}g f\,,\ \ f\in\cG^s\subset\cG:\ f(z,\bz)|_{z=0}=1
  \eq
%
leads to the finite-dimensional reduced phase space ${\rm P}^{red}$.
The reduced Poisson algebra coincides with (\ref{vv032}) for the
elliptic $r$-matrix (\ref{vv409}), and
   \beq\label{vv589}
  \begin{array}{c}
  \displaystyle{
g(z,\bz)|_{\,{\rm P}^{red}}=\mathrm
L(z)=T_0S_0+\tr_2(r_{12}(z)S_2)=T_0S_0+\sum\limits_{\al\neq 0}
\vf^0_\al(z)T_\al S_\al\,.
   }
   \end{array}
  \eq
From (\ref{gta}) it follows that
   \beq\label{vv5891}
  \begin{array}{c}
  \displaystyle{
\tr\bA\to \tr\bA+\tr (f^{-1}\bar\p f)\,.
   }
   \end{array}
  \eq
  Since
$\tr\bA$ is double-periodic it can be gauge transformed
  to a constant $a$, i.e.
   \beq\label{vv5892}
  \begin{array}{c}
  \displaystyle{
\tr\bA(z,\bz)|_{\,{\rm P}^{red}}=a\,.
   }
   \end{array}
  \eq
Using the arguments similar to (\ref{vv581})-(\ref{vv583}) one can
see that
   \beq\label{vv5893}
  \begin{array}{c}
  \displaystyle{
\{a,\mathrm L(z)\}=\mathrm L(z)\,.
   }
   \end{array}
  \eq
  and reproduce the Poincar\'e algebra (\ref{pa}) in the form of
  (\ref{vv584}) with $g=\mathrm L(z)$.
The variable $a$ extends the phase space of the top. The variable
dual to $a$ is the one which acts on the top variables by dilatation
$S\to \lambda S$. This action does not preserves the values of the
Casimir functions generated by $\det \mathrm L(z,S)$. Therefore, the
Poincar\'e symmetry emerges on the top's phase space extended by the
two-dimensional space $(a,\la)$ -- cotangent bundle to the
one-dimensional center of the group.



\subsection{Large $N$ limit: 2d elliptic hydrodynamics}

In this paragraph we consider the large $N$ limit of the elliptic
tops following and \cite{KLO,O} (see also \cite{AALOZ}), where
$\eta$-independent case was studied. This type of limit leads to 2d
hydrodynamics \cite{Ar}. The idea is to replace the generators
(\ref{vv401}) of ${\rm gl}_N$ with
 \beq\label{3.10}
T_a:=\frac{i}{2\pi\te}\,\exp \left(2\pi\imath\frac{a_1a_2}{2}\te
\right)U_1^{a_1}U_2^{a_2}\,~~a\in{\mZ}^{(2)}=\mZ\oplus\mZ\,.
 \eq
While $Q$ and $\Lambda$ from (\ref{vv401}) gives the
finite-dimensional representation of the Heisenberg group, the
generators $U_1$ and $U_2$ satisfy commutation relation of the
noncommutative torus ${\cal T}^2_\te$:
 \beq\label{3.1}
U_1U_2=e^{ -2\pi i \te} U_2U_1\,,~ \te\in[0,1)\,.
 \eq
A generic element from ${\cal T}^2_\te$ is
$X=\sum\limits_{a_1,a_2\in{\mathbb
Z}}c_{a_1,a_2}U_1^{a_1}U_2^{a_2}$, $c_{a_1,a_2}\in\mathbb C$. This
space is naturally identified with  smooth functions on the
two-dimensional torus $T^2=\{\mR^2/\mZ\oplus\mZ\}$:
 \beq\label{3.122}
U_1\to\exp(2\pi\imath x_1)\,,~U_2\to\exp(2\pi\imath x_2)\,,\ \
0<x_1,x_2\leq 1
 \eq
 with the Moyal multiplication
 \beq\label{3.123}
\exp(2\pi\imath x_1)\star\exp(2\pi\imath
x_2)=e^{-2\pi\imath\te}\exp(2\pi\imath x_2)\star\exp(2\pi\imath
x_1)\,,
 \eq
or
 \beq\label{3.3}
(f\star g)(x):=fg+
\sum_{n=1}^\infty\frac{(\imath\pi\te)^n}{(2\pi\imath)^{2n}n!}
\,\ve_{r_1 p_1}\ldots\ve_{r_n p_n}\,(\p^n_{x_{r_1}\ldots x_{r_n}}f)
(\p^n_{x_{p_1}\ldots x_{p_n}}g)
 \eq
 for functions
 \beq\label{3.2}
f(x)=\sum_{a\in\mZ^{(2)}}f_aT_a(x)\,,
~~T_a(x)=\frac{\imath}{2\pi\te}\,\exp
\left(2\pi\imath\frac{a_1a_2}{2}\te \right)\exp(2\pi\imath a_1
x_1)\exp(2\pi\imath a_2 x_2)\,.
 \eq
Then
 \beq\label{3.29}
U_1 f(x)=f(x-\theta)\,,\ \ \ U_2f(x)=\exp(2\pi\imath x) f(x)\,.
 \eq
 In other words, $U_1$ and $U_2$ are ${\rm GL}(\infty)$ analogues of
 $Q$ and $\Lambda$ from (\ref{vv401}).
 The finite-dimensional relations $T_a T_b=\kappa_{a,b}T_{a+b}$
are saved in the infinite-dimensional case with
 \beq\label{mnc}
\kappa_{a,b}^\te=-2\pi\imath\te\exp(\pi\imath\te\, a\times
b)\,,~~(a\times b= a_2b_1 -a_1b_2)\,.
 \eq
 Then we can introduce the following generalization of the Belavin's $R$-matrix (\ref{vv005})
 \cite{KLO}:
  \beq\label{BRM}
 \begin{array}{c}
  \displaystyle{
 R^\hbar_{12}(z|\,\te,\ep)=
 \sum\limits_{a\in\,{\mathbb Z}\times{\mathbb Z}} \vf_a^\hbar(z|\,\te,\ep)\, T_a\otimes
 T_{-a}
 }
 \end{array}
 \eq
Here
 \beq\label{vaf} \vf_a^\hbar(z|\,\te,\ep)
 =\exp(2\pi\imath\ep_2 a_2
z\te)\,\phi((\ep_1a_1+\tau \ep_2a_2)\te+\hbar,z)\,, \eq
$$
\ep=(\ep_1,\ep_2)\,,~~\ep_a\te<1\,.
$$
and $T_a$ is the basis (\ref{3.10}). It satisfies the Yang-Baxter
equation (\ref{vv002}).
The related quantum $L$-operator is similar to (\ref{vv006}):
 \beq\label{ql}
 \begin{array}{c}
  \displaystyle{
 \hat L^\eta(z|\,\te,\ep)\stackrel{(\ref{vv0025})}{=}\tr_2 \left(R^{\,\eta}_{12}(z)\hat S_2\right)=
 \sum\limits_{a\in\,{\mathbb Z}_N\times{\mathbb Z}_N} \vf_a^\eta(z|\,\te,\ep)\,
 T_a\,\hat S_a\,,
 }
 \end{array}
 \eq
where $\tr$ is defined as trace functional  on ${\cal T}^2_\te$:
 \beq\label{tr0}
\langle X\rangle= \tr(X)=c_{00}\,, \ \ \langle
1\rangle=1\,,~~~\langle XY\rangle=\langle YX\rangle\,.
 \eq
In the Moyal representation: $\tr
f=-\f1{4\pi^2}\int_{\cT^2_\te}fdx_1dx_2=f_{00}$.

 The Lax operator (\ref{ql}) satisfies the quasi-periodicity conditions
  \beq\label{qp}
\left\{
 \begin{array}{l}
L^\eta(z+1)=U_1^{-\ep_2}L^\eta(z)U_1^{\,\ep_2}\,,  \\ \ \\
L^\eta(z+\tau)=\exp(2\pi\imath\eta)U_2^{-\ep_1}
L^\eta(z)U_2^{\,\ep_1}\,,
 \end{array}
\right.
 \eq
Here $\epsilon_1$ and $\epsilon_2$ are arbitrary real numbers in the
sense of (\ref{3.29}). Conditions (\ref{qp}) mean that $L^\eta(z)$
is a section of the (twisted) Higgs bundle over elliptic curve
$\Si_\tau$
 with the structure group $SIN_\te$. The latter consists of
 invertible elements of $\mathcal T^2_\te$
 (see e.g. \cite{AALOZ}).
The exchange relations (\ref{vv001}) provide the direct
generalization
of the Sklyanin algebra (\ref{vv007}): 
 \beq\label{ncsa}
\begin{array}{c}
  \displaystyle{
{\mathcal A}_{\tau,\hbar,\eta,\te,\ep}^{\hbox{\tiny{Skl}}}:
 }
 \\ \ \\
  \displaystyle{
P_{12}\,L^{\eta,(0)}(\te,\ep)(\hat S)_1\, {\hat
S}_2+R^{\hbar,(0)}_{12}(\te,\ep)\, {\hat S}_1\, {\hat S}_2={\hat
S}_2\,L^{\eta,(0)}(\te,\ep)(\hat S)_1\, P_{12}+ {\hat S}_2\, {\hat
S}_1\, R^{\hbar,(0)}_{12}(\te,\ep)\,.
 }
 \end{array}
 \eq
with
 \beq\label{r0}
 \begin{array}{c}
  \displaystyle{
R^{\hbar,(0)}_{12}(z|\,\te,\ep)=\sum\limits_{a\in\,{\mathbb
Z}\times{\mathbb Z}} E_1((\ep_1a_1+\ep_2a_2\tau)\te+\hbar)\,
T_a\otimes
 T_{-a}\,,
 }
 \end{array}
 \eq
and
 \beq\label{lo}
 \begin{array}{c}
  \displaystyle{
 \hat L^{\eta,(0)}(z|\,\te,\ep)=
 \sum\limits_{a\in\,{\mathbb Z}\times{\mathbb Z}} E_1((\ep_1a_1+\ep_2a_2\tau)\te+\eta)\,
 T_a\,\hat S_a\,.
 }
 \end{array}
 \eq
Notice that this algebra depends on five parameters: $\hbar$,
$\eta$, $\tau$, $\theta\epsilon_1$, $\theta\epsilon_2$.

\vskip3mm

\noindent {\bf Classical limit} $\hbar\to 0$ leads to 2d elliptic
hydrodynamics written in the Euler-Arnold form \cite{Ar}.
First, mention that the parameters $\epsilon_{1,2}$  (\ref{vaf})
allow to define the complex structure on the noncommutative torus
${\cal T}^2_\te$. For element $X=\sum_a c_aT_a$ define
 \beq\label{cs}
\bar\p_{\ep,\tau} X=\sum_a (\ep_1a_1+\ep_2 a_2\tau)c_aT_a\,.
 \eq
The operator  $J^\eta$ (\ref{vv413}) acts as the following
pseudo-differential operator:
 \beq\label{J}
J^\eta (S)(x)=E_1(\eta+\te\bar\p_{\ep,\tau}) S(x)\,.
 \eq
Consider analogue of the classical finite-dimensional Lax pair
(\ref{vv410}), (\ref{vv411}):
 \beq\label{4.98}
 \begin{array}{c}
   \displaystyle{
  L^{\eta}(z,S(x))=\sum_{a\in\mZ^{(2)}} S_{a} \vf_a^\eta(z|\,\te,\ep)
  T_a\,,\ \ \
  M(z,S(x))=\sum_{a\in\mZ^{(2)}} S_{a} \vf_a^0(z|\,\te,\ep)
  T_a\,.}
 \end{array}
 \eq
Then the Lax equations provide equations of motion:
 \beq\label{4.4}
\p_t S(x)=\ad^*_{J^\eta(S)(x)}S(x)=[S(x),J^\eta(S)(x)]_\te\,,
 \eq
 where
 \beq\label{4.477}
 [f(x),g(x)]_\te=\te^{-1} (f\star g- g\star f)\,.
 \eq
In components we have:
 \beq\label{4.24}
  \begin{array}{c}
   \displaystyle{
  \p_tS_\al = \sum
_{\ga\in{\mZ}^{(2)}} C_\te(\al,\ga)S_\ga
S_{\al-\ga}\,E_1((\ep_1\ga_1+\ep_2\ga_2\tau)\te+\eta)\,, }
\\
   \displaystyle{
C_\te(\al,\be)=\frac{1}{\pi\te}\sin(\pi\te(\al\times\be))\,.}
  \end{array}\eq
 The obtained equations (\ref{4.4}) can be treated as hydrodynamical limit of
 the elliptic spin Ruijsenaars-Schneider model. It is an interesting problem to find
 its relation to another  type of hydrodynamical limit \cite{AW}. The
 latter approach leads to the quantum ${\rm gl}_N$ Benjamin-Ono and
 KdV systems while our approach gives rise to the ${\rm gl}_N$ Sklyanin
 type
 algebra (\ref{ncsa}) and classical equations (\ref{4.4}), (\ref{4.477}).

In the {\em non-relativistic limit} $\eta\to 0$ we reproduce the
answer from \cite{KLO,O}. Likewise the RS model goes into CM one,
the relativistic top goes to non-relativistic (\ref{vv415}). In the
same way instead of $J^\eta$ (\ref{J}) we get
 \beq\label{J2}
\mathrm J (S)(x)=-E_2(\te\bar\p_{\ep,\tau}) S(x)\,.
 \eq

In a similar way one can describe the dispersionless limit $\te\to
0$, $\ep_{1,2}\to\infty$, $\lim_{\te\to
0}(\te\ep_{1,2})=\ep'_{1,2}<1$. In the limit the Lie algebra
$sin_\te$ of the group $SIN_\te$ becomes the Lie algebra $Ham(T^2)$
of Hamiltonian vector fields on the two-dimensional torus.

\newpage

\section{Conclusion}
\setcounter{equation}{0}

Let us briefly summarize the obtained results.

 \begin{theor}\label{theor1}
Let $
 {\hat L^\eta}(z)=\tr_2 \left(R^{\,\eta}_{12}(z)\hat S_2\right)$,
 $\hat S=\res\limits_{z=0} \hat L(z)$
be ${\rm gl}_N$ solution of the quantum exchange relations
(\ref{vv001}) with the quantum  non-dynamical $R$-matrix satisfying
(\ref{vv002}), (\ref{vv00271}) and (\ref{vv0011}), (\ref{vv003}).
Then

\noindent { 1)} The quantum exchange relations (\ref{vv001}) define
the following ${\rm gl}_N$ Sklyanin algebra:
 \beq\label{vv00749}
 \begin{array}{c}
  \displaystyle{
{\mathcal A}_{\hbar,\eta}^{\hbox{\tiny{Skl}}}:\ \ \
P_{12}\,L^{\eta,(0)}(\hat S)_1\, {\hat S}_2+R^{\hbar,(0)}_{12}\,
{\hat S}_1\, {\hat S}_2={\hat S}_2\,L^{\eta,(0)}(\hat S)_1\, P_{12}+
{\hat S}_2\, {\hat S}_1\, R^{\hbar,(0)}_{12}\,.
 }
 \end{array}
 \eq

\noindent { 2)} The
 ${\rm gl}_N$-valued Lax matrix
 \beq\label{vv47026}
  \begin{array}{c}
  \displaystyle{
 { L^\eta}(z)=\tr_2 \left(R^{\,\eta}_{12}(z) S_2\right)\,,\
 \  S=\res\limits_{z=0}  L(z)
 }
 \end{array}
 \eq
defines the classical integrable system described by the Poisson
structure
 \beq\label{vv40084}
 \begin{array}{c}
  \displaystyle{
\{L^\eta_1(z)\,, L^\eta_2(w)\}=[ L^\eta_1(z)\,
L^\eta_2(w),r_{12}(z-w)]\,,
 }
 \\ \ \\
  \displaystyle{
{\mathcal A}_{\hbar=0,\eta}^{\hbox{\tiny{Skl}}}:\ \ \
\{S_1,S_2\}=[S_1S_2,r_{12}^{(0)}]+[L^{\eta,(0)}(S)_1\,S_2,P_{12}]\,.
 }
 \end{array}
 \eq
\noindent { 3)} The simplest Hamiltonian $\tr (S)$ generates
top-like equations of motion
  \beq\label{vv40142}
  \begin{array}{c}
  \displaystyle{
\dot S=[S,J^\eta(S)]
 }
 \end{array}
 \eq
with the inverse inertia tensor
   \beq\label{vv40143}
  \begin{array}{c}
  \displaystyle{
 J^\eta(S)=\tr_2\left(\left(R_{12}^{\eta,(0)}-r_{12}^{(0)}\right)S_2\right)\,,
   }
   \end{array}
  \eq
where $R_{12}^{\eta,(0)}$ and $r_{12}^{(0)}$ are the coefficients of
the local expansion of the quantum $R$-matrix (\ref{vv0011}) and the
classical $r$-matrix (\ref{vv0031}) respectively.

\noindent { 4)} Equations (\ref{vv40142}) are presented in the Lax
form $\dot L^\eta(z)=[L^\eta(z),M(z)]$ with the $M$-operator given
in terms of the classical $r$-matrix:
 \beq\label{vv4016}
 \begin{array}{c}
  \displaystyle{
M(z)=-\tr_2\left(r_{12}(z)S_2\right)\,.
 }
 \end{array}
 \eq
\noindent { 5)} Alternatively, the relativistic top can be described
in $\eta$-independent form as bihamiltonian system with the
quadratic Poisson structure described in Section \ref{quad2} and the
linear Poisson structure given  in Section \ref{lin}. The relation
between $\eta$-dependent and $\eta$-independent descriptions are
given by (\ref{vv03149}).
 \end{theor}

\vskip3mm

 \begin{theor}\label{theor2}
The classical Lax matrix (\ref{vv352})-(\ref{vv356})  provides an
example of relativistic integrable top described in Theorem
\ref{theor1}. If $S$ is of rank one and $\det
L^\eta(z)=\frac{z+\eta}{z}$ then the model is gauge equivalent to
the rational ${\rm sl}_N$ RS model defined by the Lax matrix
(\ref{v061}) with the change of variables
 \beq\label{vv4515}
  \begin{array}{c}
 \displaystyle{
S_{ij}(\bfp,\bfq)=\sum_{m=1}^{N}\,\frac{({
q}_{m}+\eta)^{\,\varrho(i)} e^{p_{m}/c} }{ \prod\limits_{k\neq
m}^{\,} ({ q}_{m}-{
q}_{k})}\,\,\,(-1)^{\varrho(j)}\,\sigma_{\varrho(j)}(\bfq)\,.
 }
 \end{array}
 \eq
 \end{theor}

\vskip3mm

 \begin{theor}\label{theor3}
The quantum $R$-matrix (\ref{vv35222})-(\ref{vv35622}) is a unitary
solution of the Yang-Baxter equation (\ref{vv002}) with
$f^\hbar(z)=\hbar^{-2}-z^{-2}$ from (\ref{vv00271}). It can be
obtained from the Lax matrix (\ref{vv352})-(\ref{vv356}) as
 \beq\label{vv4222}
 \begin{array}{c}
  \displaystyle{
 R^\hbar_{12}(z)=\sum\limits_{k,l=1}^N\frac{\p {L}^{\hbar}(z) }{\p S_{kl}}\otimes {\mathrm
 E}_{lk}\,.
 }
 \end{array}
 \eq
 In the limit (\ref{vv7140})-(\ref{vv71400}) it gives the XXX
 $R$-matrix. In ${\rm gl}_2$ case it is the (11-vertex) Cherednik's
 $R$-matrix \cite{Cherednik}.
 \end{theor}
 The latter statement can be obtained by the direct IRF-Vertex
 transformation starting from the quantum $R$-matrix for the
 rational RS model. We will give this proof elsewhere.

\vskip1mm


The obtained rational $R$-matrix allows to define new type of spin
chains and Gaudin models. In the elliptic case, in addition to the
relativistic top  we describe the large $N$ limit as the elliptic
hydrodynamics.

\subsection*{Remarks}

\begin{itemize}


\item Relations
between the Ruijsenaars--Schneider (RS) systems and quantum
integrable chains appeared recently in the context of the
Quantum-Classical duality using the Bethe ansatz approach \cite{GZZ}
or the $\tau$-function approach \cite{Anton}. This duality, in
particular, implies the substitution $\eta=\hbar$ into the Lax
matrix of the RS model and provides an alternative (to the algebraic
Bethe ansatz) method for computation of  spectrum of the quantum
spin chains transfer-matrices.

The phenomenon of the Quantum-Classical duality type was also
observed at the level of gauge theories in the series of papers
\cite{NS}. In this approach the Planck constant (in quantum
integrable system) was identified with the twisted mass parameter in
the $\mathcal N=2^*$ SUSY Yang-Mills theory. The latter mass
parameter is related to the action of the global $U(1)$ group  on
the adjoint chiral multiplet field. It resembles the appearance of
the $\eta$ parameter in the twisted boundary conditions for the Lax
operator (\ref{vv4055}), i.e. the twisted mass and the $\eta$ play
similar roles and can be closely related.

A relation between the classical and quantum systems arises also in
studies of the spectral duality \cite{MMRZZ}. A general statement is
that the spectral duality works in the same way both at classical
and quantum levels, or the properly defined classical and quantum
spectral curves for spin chains coincide. It also resembles the
similarity of the quantum $R$-matrices and the classical Lax
operators.

\item Expression (\ref{vv35222})-(\ref{vv35622}) for the rational $R$-matrix is
complicated. In the same time, the computations of particular
examples emerge a lot of cancellations. We hope that the answer can
be simplified. It is an interesting problem to find some elegant
form for the rational $R$-matrix of group-theoretical type.

\item The classification of integrable systems of Hitchin type on elliptic curves can be naturally made in terms of
characteristic classes of underlying Higgs bundles \cite{LOSZ1,SZ}.
In particular, it allows one to obtain intermediate solutions of the
quantum Yang-Baxter equation (between pure dynamical for RS model
and pure non-dynamical one) \cite{LOSZ4}. In the rational case we
deal with degenerated (and punctured) elliptic curve $y^2=z^3$. It
is interesting to know whether elliptic classification survives in
the rational limit.

\item In this paper we do not consider the trigonometric case.
However, it would appear reasonable that the obtained results are
valid in this case as well. The corresponding non-dynamical quantum
$R$-matrix was obtained in \cite{Zabrodin1} from the trigonometric
RS model via the IRF-Vertex transformation. It can be used for
construction of trigonometric top-like models.

\end{itemize}



 \begin{small}

 \end{small}

\end{document}